\documentclass[aps,prd,groupedaddress,twocolumn,nofootinbib,superscriptaddress]{revtex4-1}
\newcommand{\be}{\begin{eqnarray}}
\newcommand{\ee}{\end{eqnarray}}
\usepackage{fancyhdr}
\usepackage{amsmath}
\usepackage{mathrsfs}
\usepackage{float}
\usepackage{xcolor}
\usepackage{graphicx, epsfig, amssymb} 
\usepackage{amsmath, amsfonts}
\usepackage{bm} 
\usepackage{tensor}
\usepackage{soul}
\usepackage{hyperref}
\usepackage{physics}

\begin{document}

\title{Simulation of primordial black hole formation using pseudo-spectral methods}

\author{Albert Escriv\`a}
\email{albert.escriva@fqa.ub.edu}
\affiliation{Institut de Ci\`encies del Cosmos, Universitat de Barcelona, Mart\'i i Franqu\`es 1, 08028 Barcelona, Spain}
\affiliation{Departament de F\'isica Qu\`antica i Astrof\'isica, Facultat de F\'isica, Universitat de Barcelona, Mart\'i i Franqu\`es 1, 08028 Barcelona, Spain}

\begin{abstract}

In this work we have used for the first time pseudo-spectral methods to perform numerical simulations of spherically symmetric black hole formations on a Friedman-Robertson-Walker universe. With these methods, the differential equations describing the gravitational collapse are partially solved algebraically. With our publicly available code we then independently check, and confirm, previous numerical estimations of the thresholds to form primordial black holes. By using an excision technique and analytical estimations of accretion rates, we were also able to estimate the black holes mass even in the case of large deviations from the threshold. There, we confirm, with an explicit example, that the estimation of the black hole mass via the self-similar scaling law is only accurate up to $O(15\%)$, for the largest allowed mass.

\end{abstract}

\maketitle

\section{Introduction}\label{sec:intro}

Primordial Black Holes (PBHs) were first considered in \cite{hawking1,hawking2}. They could have been formed in the very early Universe due to the gravitational collapse of cosmological perturbations in the radiation epoch. Within this hypothesis PBHs can be generated as
a consequence of high non-linear peaks in the primordial distribution of density perturbations. While at Cosmic Microwave Background Radiation (CMB) scales the amplitudes of
the curvature perturbations are too small to generate a significant amount of PBHs, there is
currently no hard bound on their amplitudes at smaller scales, leaving open the possibility of
having a large fraction of the Dark Matter (DM) in the form of PBHs \cite{Carr,darkmatter1,darkmatter2,darkmatter3,darkmatter4,darkmatter5,darkmatter6,darkmatter7,darkmatter8}.

Several studies have addressed the problem of estimating PBH abundances from the power spectrum and including the effect of non-Gaussianities \cite{germaniprl,Garriga, germani-vicente, yoo, rioto, rioto2,vicente-garriga,Franciolini_2018,Luca_2019,Hayato,Sam-single,Hayato2,Bullock,Pattison_2017,Young_2014,Kawasaki,garrigavicentejudithescriva}. In \cite{germaniprl} it was proved using peak theory that abundances of PBHs generated by the inflationary power spectrum depend strongly on the shape of the peak. The abundance turned out to be exponentially sensitive to the threshold $\delta_{c}$ of PBH formation. Analytic estimates of the threshold obtained so far \cite{carr75}, \cite{harada} are too poor to be used in this respect and therefore numerical techniques are needed (although that, newly has been pointed out a new universal threshold formula which is profile dependent, and agrees with numerical simulation within a deviation of $O(2\%)$, see \cite{RGE}).

Numerical simulations of PBH formation started some time ago with ~\cite{Niemeyer1,Niemeyer2}, where $\delta_{c}$ was computed and a universal scaling law (depending only on the fluid type) for the mass of the BH was found. The scaling relation was similar to the one obtained from the gravitational collapse of a scalar field~\cite{choptuik,carsten}. Later, several works have adressed the problem of PBH formation in a Friedmann-Robertson-Walker (FRW) background \cite{sasaki,Nakama_2014,refrencia-extra-jaume,mit_pbh,Moradi_2015,musco2005,musco2007,musco2013,hawke2002,musco2009}.

In particular, in ~\cite{hawke2002,musco2009} numerical simulations were performed reproducing the scaling behavior up to very small values near the threshold in a cosmological background. The value of the scaling exponent matched with the one quoted in the literate got from a perturbative treatment \cite{renormalizationcriticalcollapse,maison}, and from asymptotically flat numerical collapse simulations \cite{coleman}.

These simulations on PBHs are based mainly on the implementation of a numerical Lagrangian hydrodynamic code with finite differences based on ~\cite{maywhite}, which solves Misner-Sharp equations~\cite{misnersharp} that describes the motion of a relativistic fluid under a curved spacetime. A known drawback is the appearance of a singularity soon after the formation of the black hole, which leads the end of the evolution. To solve this, Misner-Hernandez equations \cite{misnerhernandez} (which are basically the Misner-Sharp equations but introducing a null coordinate) are used to avoid the formation of an apparent horizon and follow the subsequent evolution to determine the value of $M_{\rm BH}$. The method is based on \cite{baumgarte}. 

\raggedbottom
Motivated by the recent perspectives on primordial black hole and the implications in cosmology, we have addressed this problem, focusing on obtaining an efficient numerical method to compute the threshold $\delta_{c}$ and estimating the PBH mass $M_{\rm BH}$. In this paper, for the first time, we simulate the gravitational collapse of curvature perturbations leading to the formation of PBHs using Misner-Sharp equations with the implementation of pseudo-spectral method technique, which has  been already used in general relativity with a great success \cite{novak,Teukolsky}. 

We have been able to compute the threshold $\delta_{c}$ up to an accuracy of $O(10^{-5})$, the results match with the ones quoted in the literature. Moreover, to avoid the breaking of the simulation due to the formation of the singularity, instead of implementing null coordinates, we have used an excision technique. The mass is then found by the use of an analytical approximation of the mass accretion asymptotic behavior. We present for the first time, the values of the black hole mass for the higher allowed values in the case of a Gaussian curvature perturbation. Here we also show a deviation from the scaling law of up to $O(15\%)$ in the higher end of PBH masses, with a maximum allowed mass for the black hole of $M_{\rm max(BH)} \approx 3.7 M_{H}$. Our publicly accessible code built with Python can be found in \url{https://sites.google.com/fqa.ub.edu/albertescriva/home}. 

\section{Misner-Sharp equations}\label{sec:sharp}
The Misner-Sharp equations \cite{misnersharp} describes the motion of a spherically symmetric relativistic fluid. The starting point is to consider an ideal fluid with energy momentum tensor $T^{\mu \nu} = (p+\rho)u^{\mu}u^{\nu}+pg^{\mu\nu}$ with the following line element:
\begin{equation}
\label{metricsharp}
ds^2 = -A(r,t)^2 dt^2+B(r,t)^2 dr^2 + R(r,t)^2 d\Omega^2,
\end{equation}
where $d\Omega^{2} = d\theta^2+\sin^2(\theta) d\phi^2$ is the line element of a 2-sphere and $R(r,t)$ is the areal radius. The components of the four velocity $u^{\mu}$ (which are equal to the unit normal vector orthogonal to the hyperspace at cosmic time $t$ $u^{\mu}=n^{\mu}$), are given by $u^{t}=1/A$ and $u^{i}=0$ for $i=r,\theta,\phi$. From now on, we will use units $G_{N}=1$.

In the Einstein field equations appear the following quantities:
\begin{align}
\frac{1}{A(r,t)}\frac{\partial R(r,t)}{\partial t} &\equiv D_{t}R \equiv U(r,t),\nonumber \\
\frac{1}{B(r,t)}\frac{\partial R(r,t)}{\partial r} &\equiv D_{r}R \equiv \Gamma(r,t),
\label{covariantR}
\end{align}

where $D_{t}$ and $D_{r}$ are the proper time and distances derivatives. $U$ is the radial component of the four-velocity associated to an Eulerian frame. It measures the radial velocity of the fluid with respect to the centre of coordinates. The Misner-Sharp mass is introduced as
\begin{equation}
M(r,t) \equiv \int_{0}^{R} 4\pi R^{2} \rho \, \left(\frac{\partial R}{\partial r}\right) dr\, ,
\end{equation}
which is related with $\Gamma$, $U$ and $R$ though the constraint:
\begin{equation}
\Gamma = \sqrt{1+U^2-\frac{2 M}{R}}.
\end{equation}
The mass $M(r,t)$ includes contributions from the kinetic energy and gravitational potential energies. Finally, the Misner-Sharp equations governing the evolution of a spherically symmetric collapse in non-linear full general relativity are:
\begin{align}
D_{t}U &= -\left[\frac{\Gamma}{(\rho+p)}D_{r}p+\frac{M}{R^{2}}+4\pi R p \right], \\
D_{t} R &= U, \\
D_{t}\rho &= -\frac{(\rho+p)}{\Gamma R^{2}}D_{r} (U R^{2}), \\
D_{t}M &= -4 \pi R^{2} U p, \\
\label{eqconstraint}
D_{r}M &= 4\pi \Gamma \rho R^{2}, \\
\label{lapse}
D_{r} A &= \frac{-A}{\rho+p}D_{r}p\, .
\end{align}

The boundary conditions are $R(r=0,t)=0$, leading to $U(r=0,t)=0$ and $M(r=0,t)=0$. Then, by spherical symmetry, we have $D_{r} p (r=0,t)=0$.

\section{Cosmological set up for PBH formation}\label{sec:cosmos}

We apply the Misner-Sharp equations in the cosmological context within a FRW background. To close the system we need to give the equation of state of the fluid, which in our context is $p=\omega \rho$. At $r \rightarrow \infty$ we want to match with the FRW background, but in a numerical simulation we have to handle with a finite grid. Then, to match the outer point of the grid with the FRW solution and to avoid reflections from pressure waves, we have used the condition $D_{r} p (r=r_f,t) = 0$ (where $r_{f}$ if the outer point of the grid). Eq.(\ref{eqconstraint}) is called the Hamiltonian constraint, we will use it later on for numerical checks. Eq.(\ref{lapse}) can be solved analytically imposing $A(r_f,t) = 1$ to match with the FRW spacetime. This gives:
\begin{equation}
A(r,t) = \left(\frac{\rho_{b}(t)}{\rho(r,t)}\right)^{\frac{\omega}{\omega+1}},
\end{equation}
where $\rho_{b}(t) = \rho_{0}(t_{0}/t)^{2}$ is the energy density of the FRW background and $\rho_{0}=3 H_{0}^{2}/8\pi$. Using the definitions of Eq.(\ref{covariantR}), we can rewrite Misner-Sharp equations in a more convenient way to perform the numerical simulations:

\begin{align}
\label{eq:msequations1}
\dot{U} &= -A\left[\frac{\omega}{1+\omega}\frac{\Gamma^2}{\rho}\frac{\rho'}{R'} + \frac{M}{R^{2}}+4\pi R \omega \rho \right], \\
\dot{R} &= A U, \\
\label{eq:msequations2}
\dot{\rho} &= -A \rho (1+w) \left(2\frac{U}{R}+\frac{U'}{R'}\right), \\
\dot{M} &= -4\pi A \omega \rho U R^{2},
\end{align}

where $(\dot{})$ and $(')$ represents the time and radial derivative respectively. At superhorizon scales the metric Eq.(\ref{metricsharp}) can be approximated, at leading order in gradient expansion, by the following metric \cite{sasaki}:

\begin{equation}
\label{frwmetric}
ds^2 = -dt^2 + a^2(t) \left[\frac{dr^2}{1-K(r) r^2}+r^2 d\Omega^2 \right].
\end{equation}

The cosmological perturbation will be encoded in the initial curvature $K(r)$. At leading order in gradient expansion, the product $K(r)r^{2}$ is proportional to the compaction function $C(r)$ \cite{sasaki}, which represents a measure of the mass excess inside a given volume. More specifically,
\begin{equation}
C(r,t) = \frac{2 \left[M(r,t)-M_{b}(r,t)\right]}{R(r,t)}.
\label{compactionfunction}
\end{equation}
We now define the location of the maximum of $C(r)$ as $r_{m}$, its value $C_{\rm max}=C(r_{\rm m})$ is going to be used as a criteria for PBH formation \cite{refrencia-extra-jaume,sasaki}. By defining $\epsilon(t) = R_{H}(t)/a(t)r_{m}$, one can solve Misner-Sharp equations at leading order in $\epsilon \ll 1$. $R_{H}(t)=1/ H(t)$ is the cosmological horizon and $r_{m}$ is the length scale of the perturbation. This approach is the so-called long wavelength approximation \cite{sasaki} (or gradient expansion). We have:

\begin{align}
\label{expansion}
A(r,t) &= 1+\epsilon^2(t) \tilde{A}(r),\nonumber\\
R(r,t) &= a(t)r(1+\epsilon^2(t) \tilde{R}(r)),\nonumber\\ 
U(r,t) &= H(t) R(r,t) (1+\epsilon^2(t) \tilde{U}(r) ),\nonumber\\ 
\rho(r,t) &= \rho_{b}(t)(1+\epsilon^2(t)\tilde{\rho}(r)),\nonumber\\ 
M(r,t) &= \frac{4\pi}{3}\rho_{b}(t) R(r,t)^3 (1+\epsilon^2(t) \tilde{M}(r) ),\nonumber\\ 
\end{align}
where for $\epsilon \rightarrow 0$ we recover the (FRW) solution. The perturbations of the tilde variables in the linear regime were computed in \cite{musco2007}, which we summarize here:
\begin{align}
\label{perturbations}
\tilde{\rho}(r) &= \frac{3(1+\omega)}{5+3\omega}\left[K(r)+\frac{r}{3}K'(r)\right] r^2_{m},  \nonumber\\
\tilde{U}(r) &= -\frac{1}{5+3\omega}K(r) r^2_{m},\nonumber\\
\tilde{A}(r) &= -\frac{\omega}{1+\omega} \tilde{\rho}(r),\nonumber\\
\tilde{M}(r) &= -3(1+\omega) \tilde{U}(r),\nonumber\\
\tilde{R}(r) &= -\frac{\omega}{(1+3\omega)(1+\omega)}\tilde{\rho}(r)+\frac{1}{1+3\omega}\tilde{U}(r).\nonumber\\
\end{align}

The background solution equations are: $H(t)=H_{0} t_{0}/t$ , $a(t)= a_{0}(t/t_{0})^{\alpha}$ and $R_{H}(t) = R_{H}(t_{0})(t/t_{0})$ where $a_{0} = a(t_{0})$ , $H_{0}=H(t_{0}) = \alpha /t_{0}$ and $R_{H}(t_{0}) = 1/H_{0}$. Moreover we define $\alpha = 2/3(1+\omega)$. We establish a time scale given by $\epsilon(t_{m}) =1$, which leads $t_m = t_{0}(a_{0} r_m/R_H(t_0))^{1/(1-\alpha)}$.

The amplitude of a cosmological perturbation can be measured by the mass excess within a spherical region:
\begin{equation}
\label{massexcess}
\delta(r,t) = \frac{1}{V}\int_{0}^{R} 4\pi R^{2} \frac{\delta \rho}{\rho_{b}} R' dr,
\end{equation}
where $V=4\pi R^{3}/3$ and at leading order in $\epsilon$ gives:
\begin{equation}
\delta(r,t)=\left(\frac{1}{a H r_{m}}\right)^{2} \bar{\delta}(r),
\end{equation}

where $\bar{\delta}(r)=f(w)K(r)r^{2}_{m}$ and  $f(\omega)= 3(1+\omega)/(5+3\omega)$. In the long wavelength approximation, $C(r,t)\simeq C(r)=f(\omega) K(r) r^{2} = r^2 \bar{\delta}(r) /r^2_{m}$ \cite{musco2018}, which yields $C(r_{m}) = \bar{\delta}(r_{m})=\bar{\delta}_{m}$. Because of the above definitions the value of $r_{m}$ is given by the solution of:
\begin{equation}
\label{cmax}
K(r_{m})+\frac{r_{m}}{2}K'(r_{m}) = 0.
\end{equation}
After the initial conditions are given the compaction function starts to evolve non-linearly and becomes time dependent. The first apparent horizon is then formed whenever the maximum of the compaction function is about one (for a more formal discussion see \cite{muscoelis}). We define the threshold for primordial black hole formation as $\delta_{c}$ such that a PBH is formed whenever $\bar{\delta}(r_{m})\geq \delta_c$. \footnote{Here we use a slightly different notation for $\delta_{m}$ from the paper of \cite{musco2018} to avoid confusion due to the use of the linear extrapolation.}

\section{Pseudo-spectral technique}\label{sec:spectral}

Instead of using a Lagrangian hydrodynamic technique with finite differences, we have implemented the Pseudo-spectral Chebyshev collocation method to compute the spatial derivatives part of the Eqs.(\ref{eq:msequations1}, \ref{eq:msequations2}). The time evolution is instead solved with fourth-order explicit Runge-Kutta method. In the following we explain the use of the pseudo-spectral technique, see also \cite{spectrallloyd} and \cite{spectralmatlab}.

Consider a function $f(x)$ and fit with $N_{\rm cheb}$ Chebyshev polynomials (although this could be any kind of orthonormal function). More specifically we can define the approximated function:
\begin{equation}
f_{N_{\rm cheb}}(x) = \sum_{k=0}^{N_{\rm cheb}} c_{k} T_{k}(x),
\end{equation}
where $T_{k}(x)$ are the Chebyshev polynomial of order $k$. The coefficients $c_{k}$, $k=0,1,...,N_{\rm cheb}$ are then obtained by solving $f_{N_{\rm cheb}}(x_k)=f(x_k)$ where $x_{k}=\cos(k \pi/N_{\rm cheb})$. Those points are called Chebyshev collocation points and correspond to $T'_{k}(x_k)=0$. 
The solution is 
\begin{align}
f_{N_{\rm cheb}}(x) &= \sum_{k=0}^{N_{\rm cheb}} L_{k}(x) f(x_{k}),\\
L_{k}(x) &= \frac{(-1)^{k+1}(1-x^2)T'_{N_{\rm cheb}}(x)}{\bar{c}_{k}N_{\rm cheb}^2(x-x_{k})} ,
\end{align}
where $\bar{c}_{k} =2$ if $k=0,N$ and $\bar{c}_{k}=1$ in other cases. The functions $L_k$ are called Lagrange interpolation polynomials. With this we can easily obtain the $p$ derivative to be:
\begin{equation}
f^{(p)}_{N_{\rm cheb}}(x_{i}) = \sum_{k=0}^{N_{\rm cheb}} L^{(p)}_{k}(x_{i}) f_{N_{\rm cheb}}(x_{k}).
\end{equation}
Defining the Chebyshev differentiation matrix $D^{(p)}=\{L^{(p)}_{k}(x_{i})\}$ we have :
\begin{align}
D^{(1)}_{i,j} &= \frac{\bar{c}_{i}}{\bar{c}_{j}}\frac{(-1)^{i+j}}{(x_{i}-x_{j})} , (i \neq j), i,j = 1,...,N_{\rm cheb}-1 ,\\
D^{(1)}_{i,i} &= -\frac{x_{i}}{2(1-x_{i}^2)} , i=1,...,N_{\rm cheb}-1,\\ 
D^{(1)}_{0,0} &=-D^{(1)}_{N_{\rm cheb},N_{\rm cheb}} = \frac{2N_{\rm cheb}^2+1}{6}\,.
\end{align}
We use the following identity to compute the diagonal terms of the matrix $D$ quoted before:
\begin{equation}
D^{(1)}_{i,i} = -\sum_{j=0,j \neq i}^{N_{\rm cheb}} D_{i,j}^{(1)},
\end{equation}
which gives a substantial improvement regarding the round-off errors in the numerical computations (see \cite{spectralmatlab} for details).

\bigskip

The crucial advantage of spectral methods in comparison with finite differences is that the error decays exponentially in $N_{\rm cheb}$. With finite differences instead, error decays like $1/N^{v}$ , where $N$ is again the sample of points and $v$ is a positive number. Moreover a crucial benefit of spectral methods respect to finite differences is that the derivative at a given point is computed globally taking into account the value of all the other points, in comparison with finite differences where the derivative at a given point only takes into account the neighbours.

In our particular case, the domain of the radial coordinate is given by $\Omega = [r_{\rm min},r_{\rm max}]$ where $r_{\rm min}=0$ and $r_{\rm max} = N_{H} R_{H}(t_{0})$. $N_{H}$ is the number of initial cosmological horizon, which in general is taken to be $N_{H} \sim 90$ as it is done in the literature \cite{musco2005}. Since our domain is not $[-1,1]$ (which is the domain for the Chebyshev polynomials), we need to perform a mapping between the spectral domain to the physical one. We have used the following linear mapping (other options are possible):
\begin{equation}
\tilde{x}_{k} = \frac{r_{max}+r_{min}}{2}+\frac{r_{max}-r_{min}}{2}x_{k}.
\end{equation}
$\tilde{x}_{k}$ are the new Chebyshev points rescaled to our domain $\Omega$. In the same way, the Chebyshev matrix can be rescaled in a straightforward way using the chain rule:
\begin{equation}
\tilde{D} = \frac{2}{r_{\rm max}-r_{\rm min}} D.
\end{equation}

To implement a Dirichlet boundary condition at given $x_{k}$, such that $f(x=x_{k})=u_{D,bc}$, it is only needed to fulfil $f_{N_{\rm cheb}}(x=x_{k})=u_{D,bc}$. Instead, in case of Neumann boundary condition such that $f^{(1)}(x=x_{k})=u_{N,bc}$, then $(D \cdot f_{N_{\rm cheb}})(x=x_{k})=u_{N,bc}$. The stability of the method depends on the value of $N_{\rm cheb}$ and $dt$ used. An increment of the spatial resolution will require an enough small time step $dt$ to avoid instabilities during the evolution.

\section{Numerical procedure}

In this section and in the rest of the paper we will test our code in a radiation dominated universe, because of its interest in PBH formation. In other words, we will fix $\omega=1/3$ and therefore $f(\omega) = 2/3$. In all our numerical simulations we are setting $t_{0}=1$ and $a_{0}=1$, which yields $H_{0}=1/2$, $R_{H}(t_{0})=2$. For the length scale of the perturbation, we have taken $r_{m}=10 R_{H}(t_{0})$ as done in the literature \cite{musco2007}, giving $t_m=10^{2} t_{0}$. This ensures that the long wavelength approximation is fulfilled. To find $\delta_{c}$ we have implemented a bisection method which scans different regimes of $\bar{\delta}$ until finding the range in which the collapse will happen. The threshold $\delta_{c}$ is defined as the mid point of this range. 

It's useful to know that $\delta_{c}$ is bounded from above by $\delta_{c} =f(\omega)$. This can be directly inferred by noticing that since $\Gamma^2 = 1-K(r)r^2$, then $K(r_m)r^2_m=1$ as maximum. The numerical procedure that we have established is described as follows:
\begin{itemize}
\item  Set up the number of Chebyshev points $N_{\rm cheb}$ and create the grid of points $x_{k}$. This yields the Chebyshev differentiation matrix $D$.
\item Introduce the initial time step $dt_{0}$ and the length scale value $r_{m}$.
\item Choose a lower and an upper bound in $\bar{\delta}$ to perform the domain of the bisection method. In our case, we have chosen $\delta_{\rm max}=2/3$ and $\delta_{\rm min} =2/5$ \cite{RGE} (although this can be changed to establish a domain closer to $\delta_{c}$ to reduce the computational time). 
\item Given a curvature profile $K(r)$ , such that $K(r)=\mathcal{A} \bar{K}(r)$ with $\bar{K}(0)=1$, compute the tilde perturbations in the other hydrodynamical magnitudes following Eqs.(\ref{expansion},\ref{perturbations}), except by the curvature amplitude $\mathcal{A}$ that multiplies all this perturbations.
\item Once the bisection method starts and a value of $\bar{\delta}_{m}$ is taken, the corresponding value of $\mathcal{A}$ is computed to set up the profile $K(r)$.
\item Use the four-order Runge-Kutta equations to integrate the equations at each time-step $dt$, imposing as well boundary conditions at each internal time step.
\item Compute at each iteration time the value of the maximum of the compaction function $C_{\rm max}$. Once it approaches $C_{\rm max} \approx 1$ an apparent horizon is formed. This corresponds to a given value of $\delta_{c,\rm yes}$ (a black hole will form). Next step is search for a lower value of $\bar{\delta}_{m}$ via bisection method modifying the bound such that $\delta_{c} \in [\delta_{\rm min} , \delta_{c, \rm yes}]$ and we go to the next iteration in the bisection. Otherwise, if $C_{\rm max} \approx C_{\rm min}$ (in our simulations we take in general $C_{\rm min}\approx 0.3$, this is related to the fact that $\delta_{\rm min} =2/5$) then the perturbation disperses (it is not going to form a black hole) getting a value $\delta_{c,\rm no}$ and we go to the next iteration in the bisection, modifying the bound such that $\delta_{c} \in [\delta_{c,\rm no}  ,  \delta_{\rm max}]$.
\item With the previous result, the bisection method is iterated until the difference between $\delta_{c,\rm yes}$ and $\delta_{c,\rm no}$ becomes less than the resolution that we set to compute the value of $\delta_{c}$, $\delta_{c,\rm yes}-\delta_{c,\rm no} \lesssim \delta(\delta_{c})$. Where we infer that $\delta_{c} = (\delta_{c,\rm yes}+\delta_{c,\rm no})/2 \pm \delta(\delta_{c})$. If during the bisection $(\delta-\delta_{c})$ goes beyond the resolution of the method, then the trial $\delta$ is shifted according to $\delta(\delta_{c})$.
\end{itemize}

For the Runge-Kutta we have used a conformal time step $dt = dt_{0}(t/t_{0})^{\alpha}$ as it improves significantly the running time. To test our code, we use the $2$-norm of the Hamiltonian constraint equation Eq.(\ref{eqconstraint}) in all the simulations, which is expected to remain constant from the beginning if Einstein equations are correctly solved during the simulations. Specifically:
\begin{align}
\mathcal{H} &= D_{r}M - 4\pi \Gamma \rho R^{2}, \\
\mid\mid\mathcal{H}\mid\mid_{2} &\equiv \frac{1}{N_{\rm cheb}}\sqrt{\sum_{k} \mid \mathcal{H}_{k} \mid^2}.
\label{eq:constraint}
\end{align}

The maximal resolution that we have been able to obtain is $\delta_{c,\rm yes}-\delta_{c,\rm no} > O(10^{-5})$. The reason is that large pressure gradients develop once $\delta$ approaches the self-similar critical solution, and so there the accuracy in computing derivatives is limited. The situation depends on the profile considered and it was already observed in \cite{musco2005}.

\section{Numerical results}

\subsection{FRW solution}
Here we check that our code reproduces the FRW solution. To do that, we have computed the relative error of the different variables $\rho,U,M,R$ ( $A$ and $\Gamma$ depends on the previous ones) with respect to the FRW analytical solution. We define $\delta X_{i} =  X(x_{i})-X_{ \rm b}(x_{i})$, where $X$ are the variables that we solve in the Misner-Sharp equations. To test our code against the FRW solution we compute the variance,
\begin{equation}
\Vert \delta X \Vert_{2} = \frac{1}{N_{\rm cheb}}\sqrt{\sum_{k} \mid \delta X_{k} \mid^2}.
\end{equation}

\begin{figure}[ht!]
\centering
\includegraphics[width=1.0\linewidth]{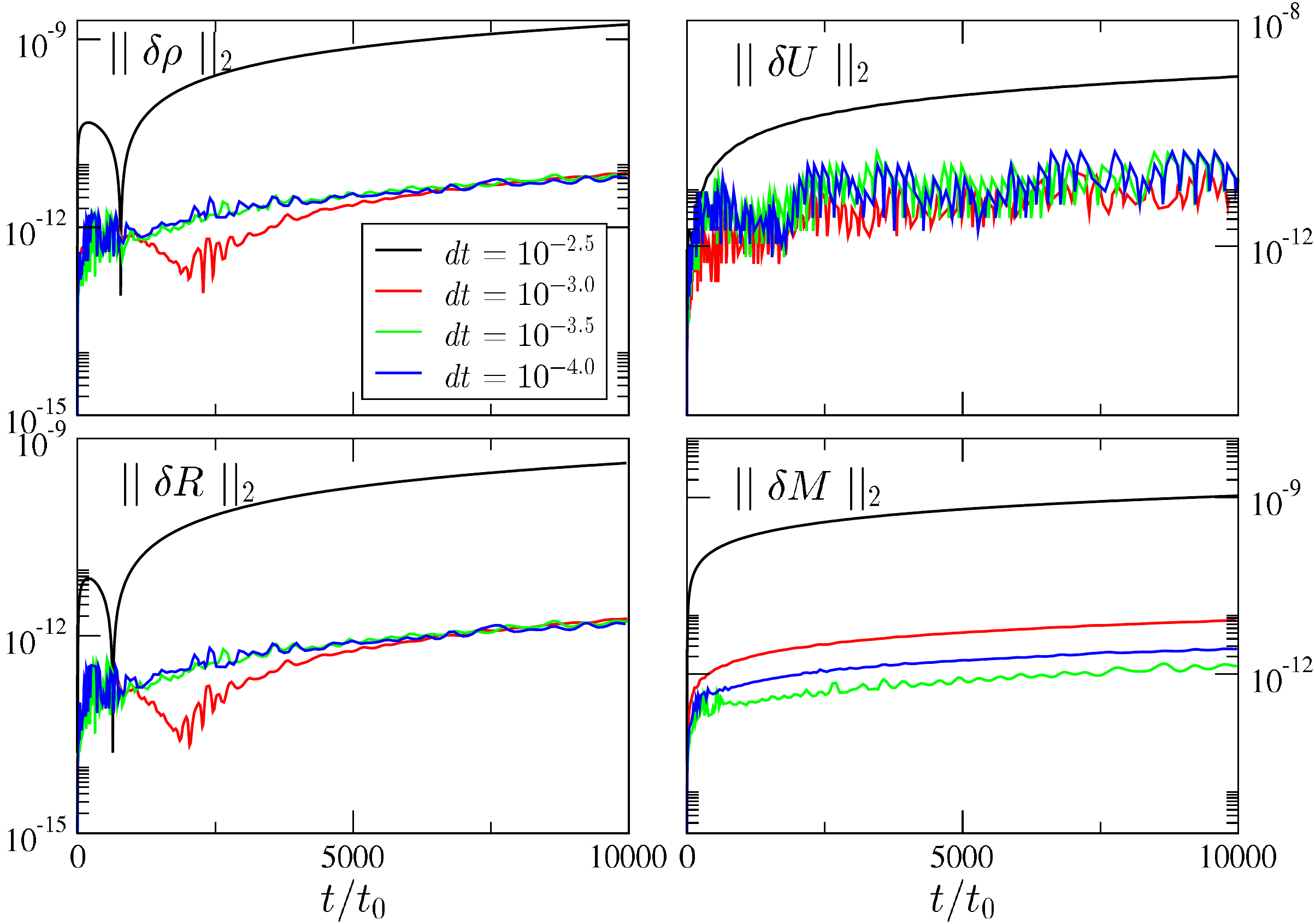} 
\caption{$\Vert \delta X \Vert_{2}$ with $N_{\rm cheb}=7$ in all cases for $dt=10^{-2.5}$ (black), $dt=10^{-3.0}$ (red), $dt=10^{-3.5}$ (green) and $dt=10^{-4.0}$ (blue).}
\label{frwNfijo}
\end{figure}

In Fig.~\ref{frwNfijo} we see $\Vert \delta X \Vert_{2}$ for the different hydrodynamical variables and we see a good convergence to the analytical solution. Already for $N_{\rm cheb}=7$ we have at least a $O(10^{-9})$ accuracy. Obviously for a curvature profile that is not homogeneous the number of Chebyshev points would need to be increased because the pressure gradients are not vanishing.

\subsection{Curvature profiles}

In this section we are going to test our code against the results obtained in \cite{musco2018} for centrally peaked profile, the ones relevant for cosmology \cite{germaniprl,germani-vicente}. In other words we shall consider the following profiles for initial curvature perturbations:
\begin{equation}
\label{profile}
\bar{K}(r) = e^{-\frac{1}{q}\left(r/r_{m}\right)^{2q}},
\end{equation}
where $q$ parametrizes the slope of the profiles. 

For $q=1$ we recover the Gaussian curvature profile. Here we get $\delta_{c} \approx 0.49774 \pm 2 \cdot 10^{-5}$, which matches the one quoted in the literature ($\delta_{c} \approx 0.5$ \cite{musco2018}). This value was obtained by using $dt_{0}=10^{-3}$ and $N_{\rm cheb}=400$. We have cheeked that this result is stable under the increment of $N_{\rm cheb}$ and/or the reduction of $dt_{0}$.

\begin{figure*}[ht!]                                                                   
\begin{center}                                                                   
\includegraphics[width=1.0\columnwidth]{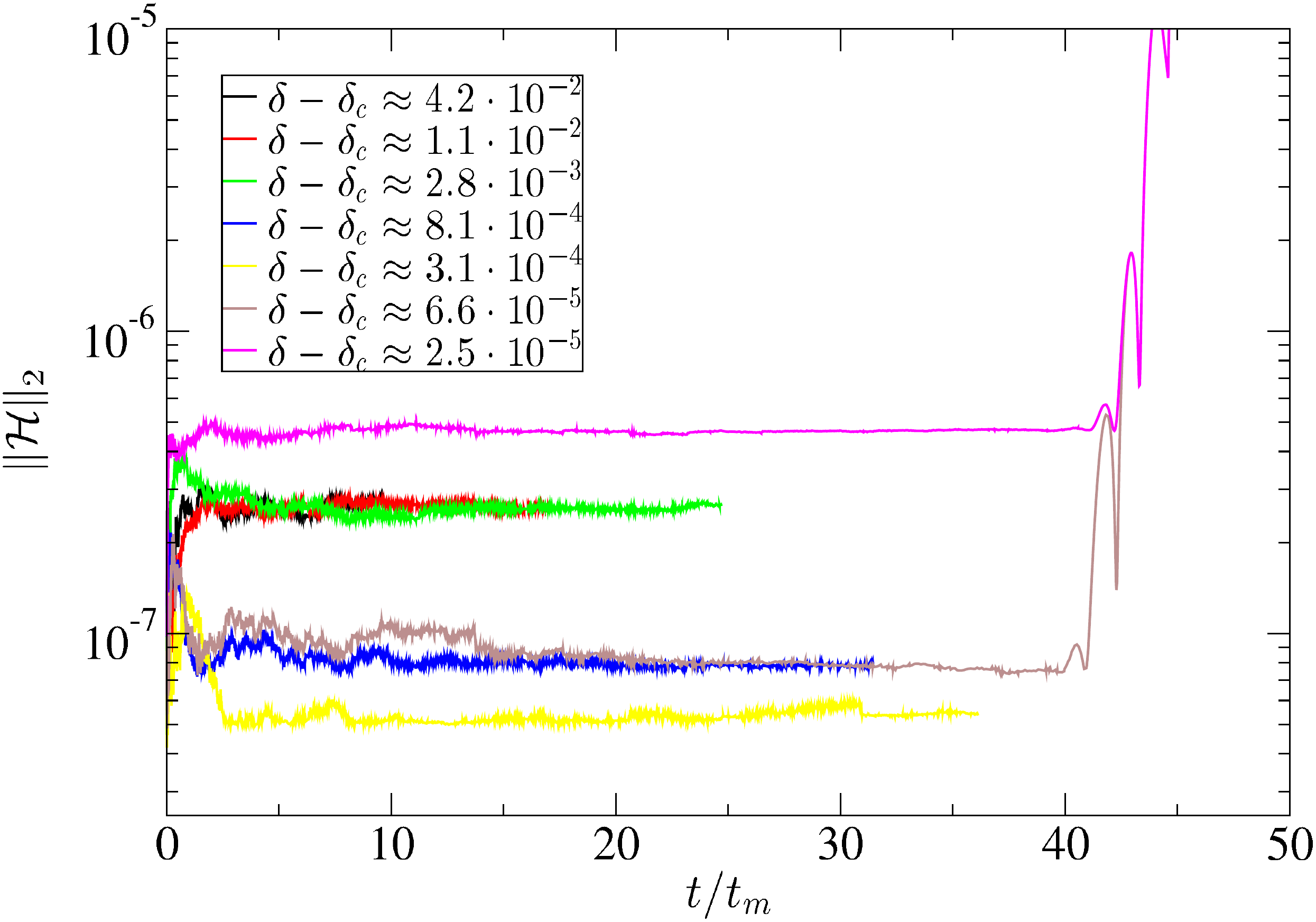}
\includegraphics[width=1.0\columnwidth]{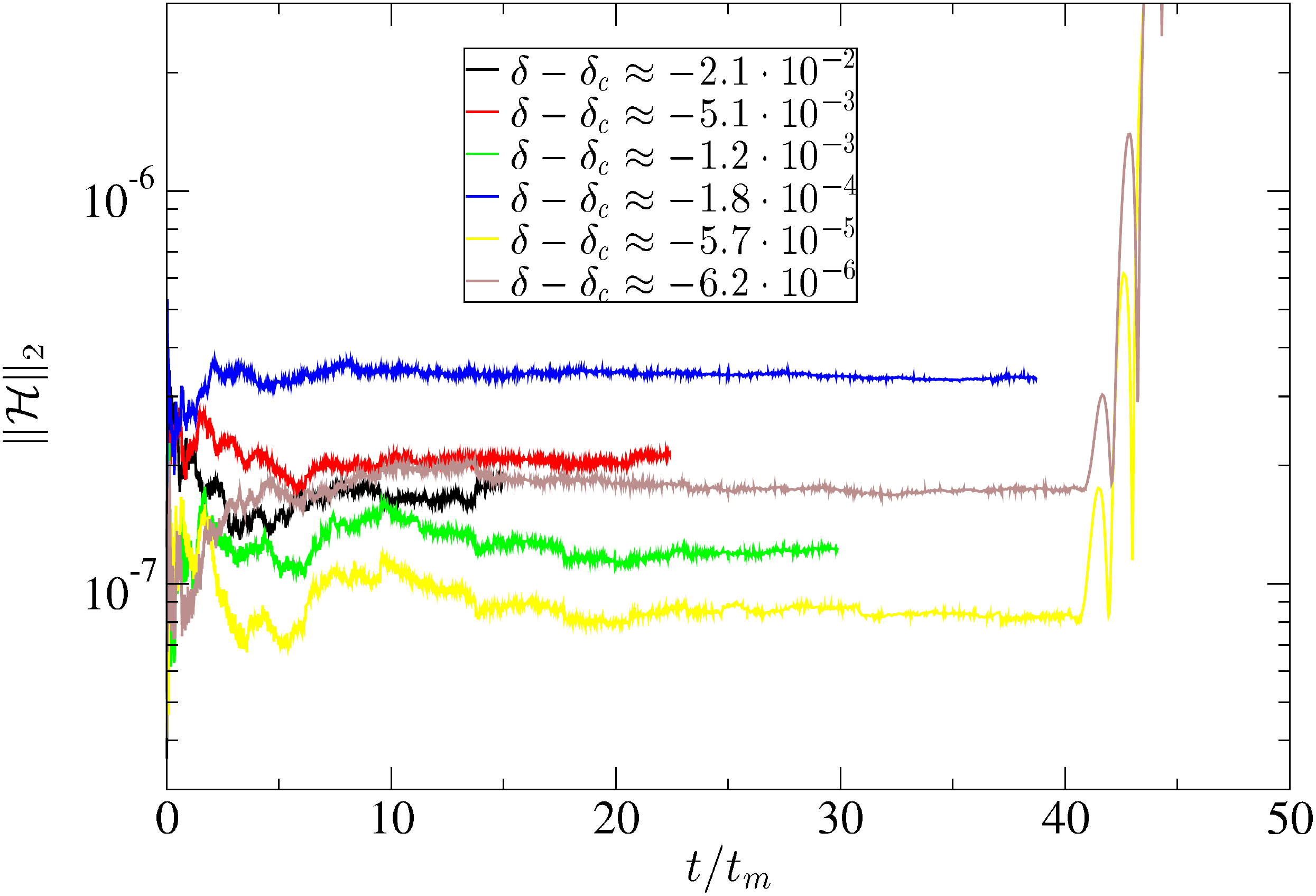}                            
\caption{
Left Panel: Hamiltonian constraint for the iterations of the bisection procedure in the case of the Gaussian curvature profile whose are leading to the formation of a black hole. Right panel: Hamiltonian constraint for the iterations of the bisection procedure in the case of the Gaussian curvature profile whose perturbations are going to disperse and not form a black hole In both cases $dt_{0} =10^{-3}$, $N_{\rm cheb}=400$. We have subtracted the initial Hamiltonian constraint for each evolution of $\delta$ in both cases.}  
\label{fig:constraintbisection}                                                                 
\end{center}                                                                     
\end{figure*}

In addition, to check the correctness of the numerical procedure of the bisection at each iteration, we have computed $\Vert\mathcal{H}\Vert_{2}$, which can be found in Fig.~\ref{fig:constraintbisection}. We see that the constraint is violated at late times for $(\delta-\delta_{c}) \approx O(10^{-5})$. This sets the maximal resolution we can achieve in this case.

\begin{figure}[ht!]
\centering
\includegraphics[width=1.0\linewidth]{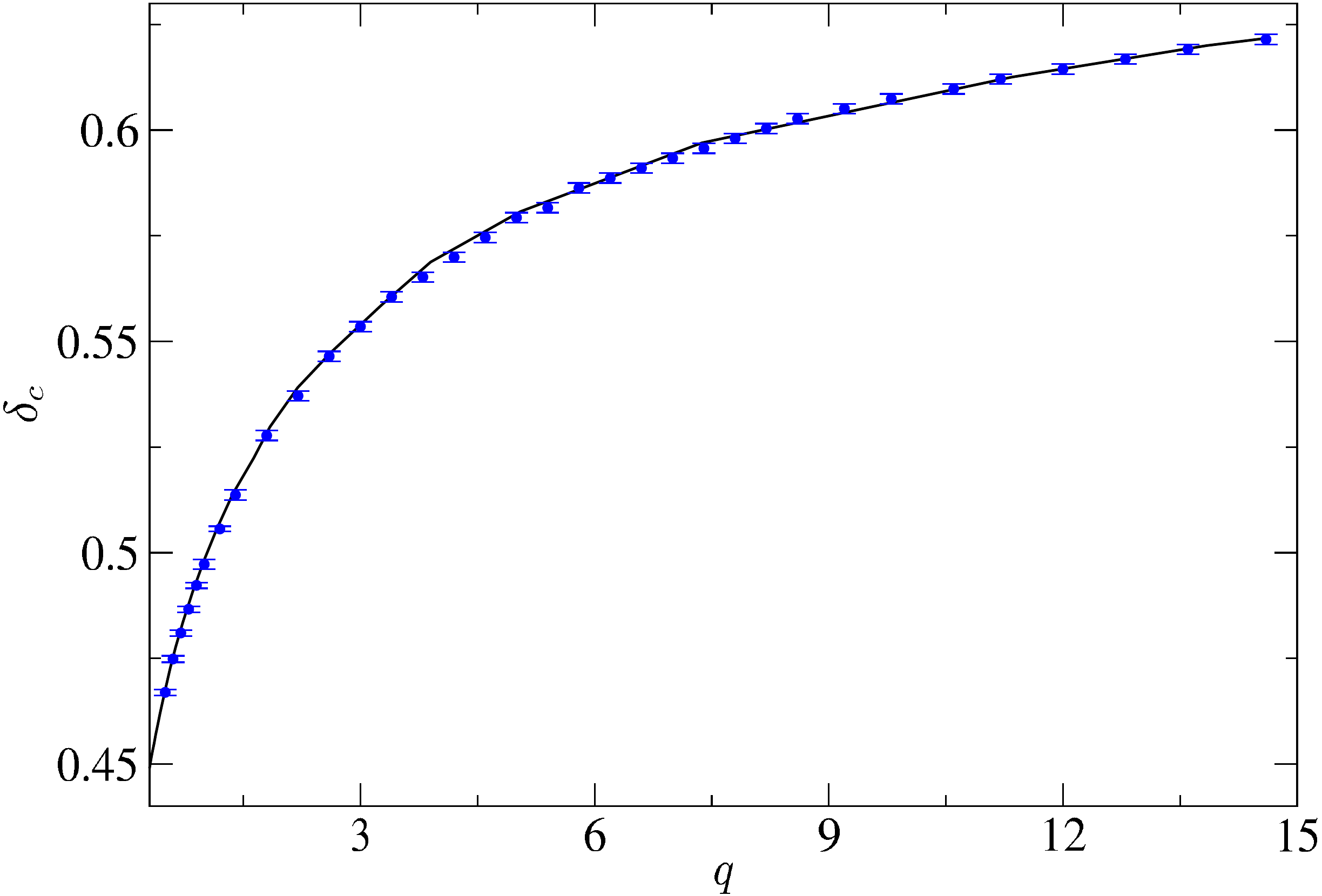} 
\caption{Values of $\delta_{c}$ for different values of $q$. Points are the values that we have got numerically. Blue line is the curve from \cite{musco2018} got using \cite{WebPlotDigitizer}. All the computations has been done with $dt_{0} =10^{-3}$ with $N_{\rm cheb}=400$, unless in some cases has been necessary to increase $N_{\rm cheb}$ to get the same accuracy in the determination of $\delta_{c}$.}
\label{fig:Iliaprofile}
\end{figure}

Finally, in Fig.~\ref{fig:Iliaprofile}, we have tested our code against the different profiles parameterized by $q$ in the range $q \in [0.5,14.6]$. Our results match with very good accuracy the ones of \cite{musco2018}.

\subsection{Gaussian profile in details}

\begin{figure*}[ht!]                                                                   
\begin{center}                                                                   
\includegraphics[width=0.8\columnwidth]{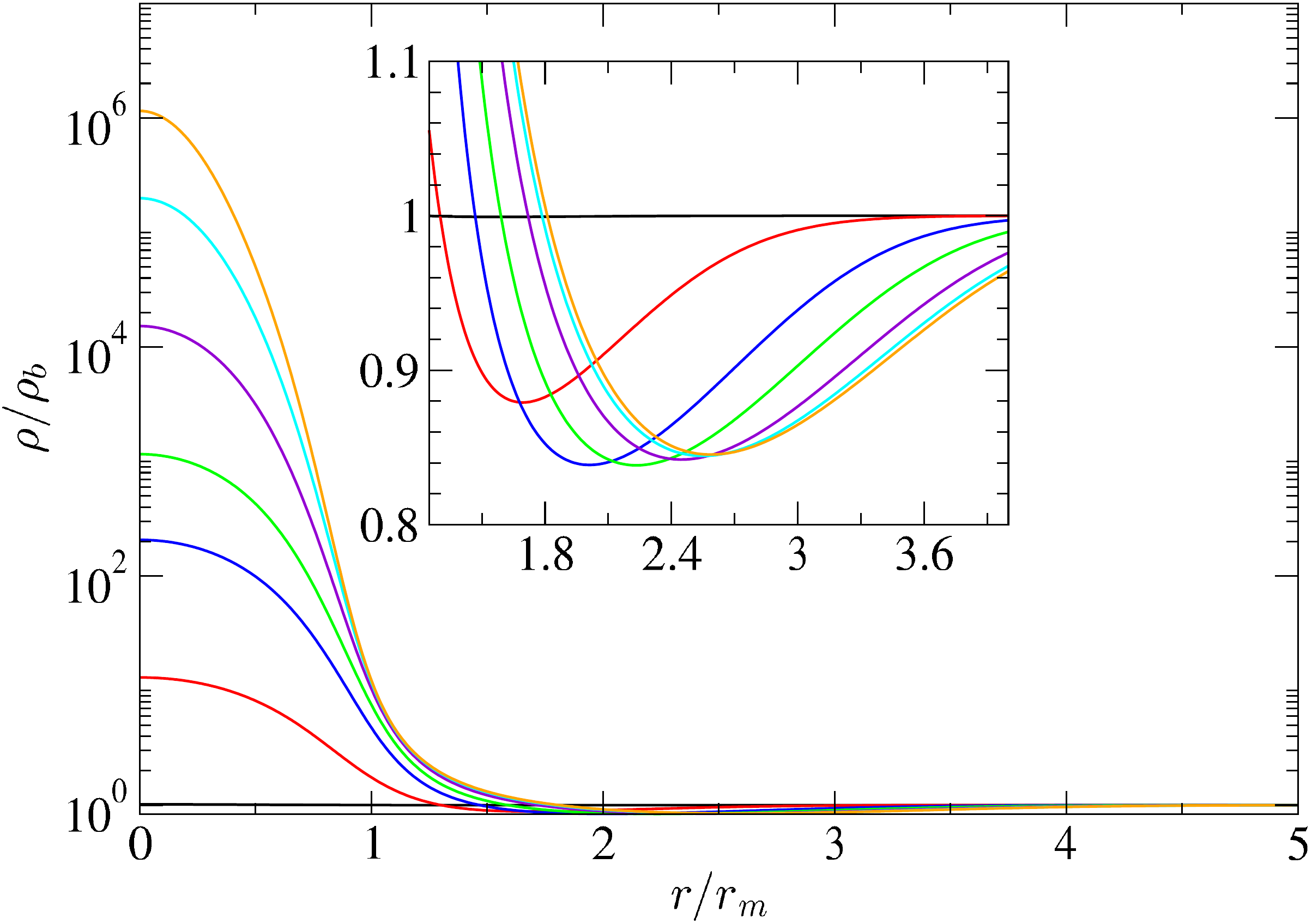}
\includegraphics[width=0.8\columnwidth]{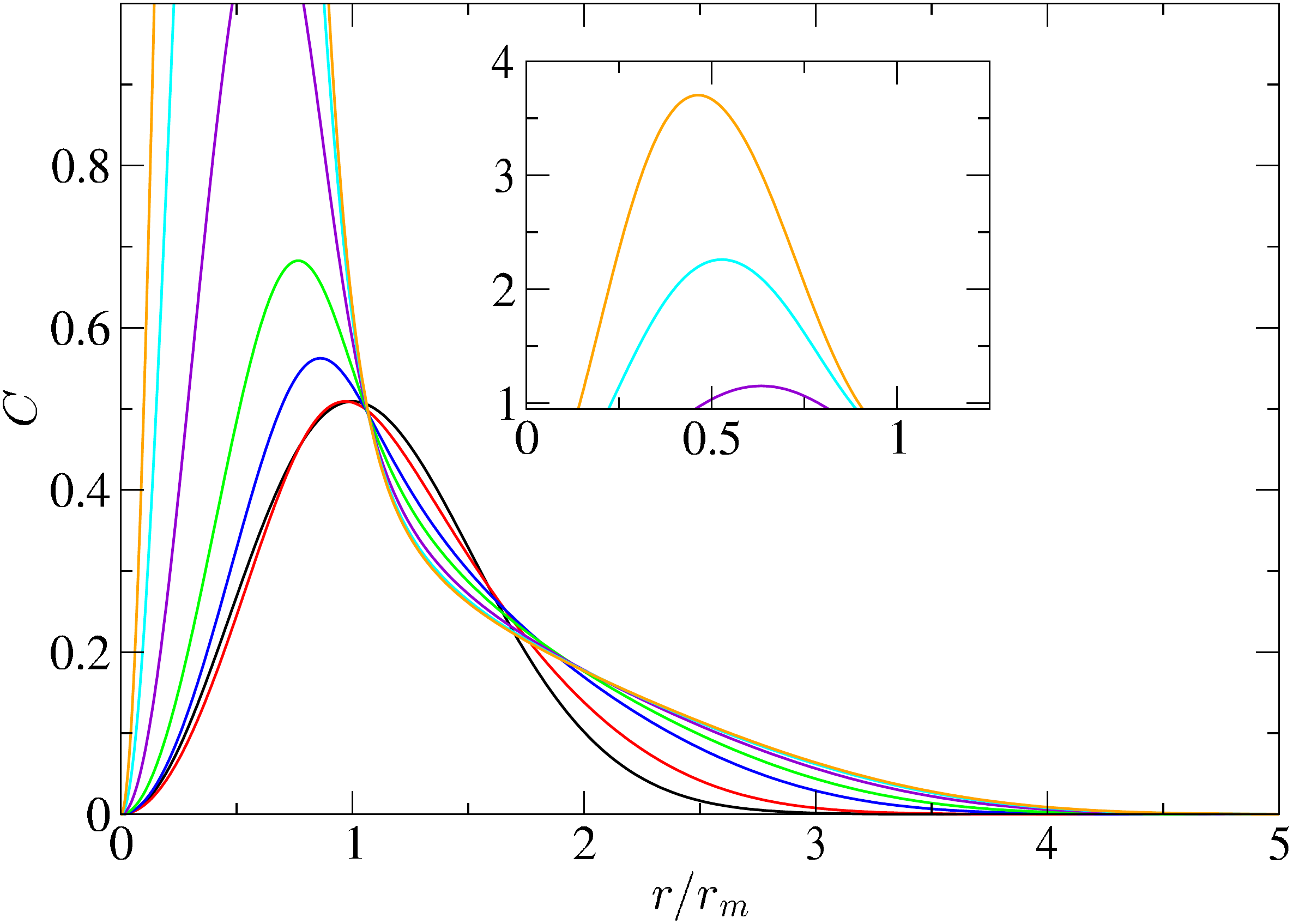}                            
\includegraphics[width=0.8\columnwidth]{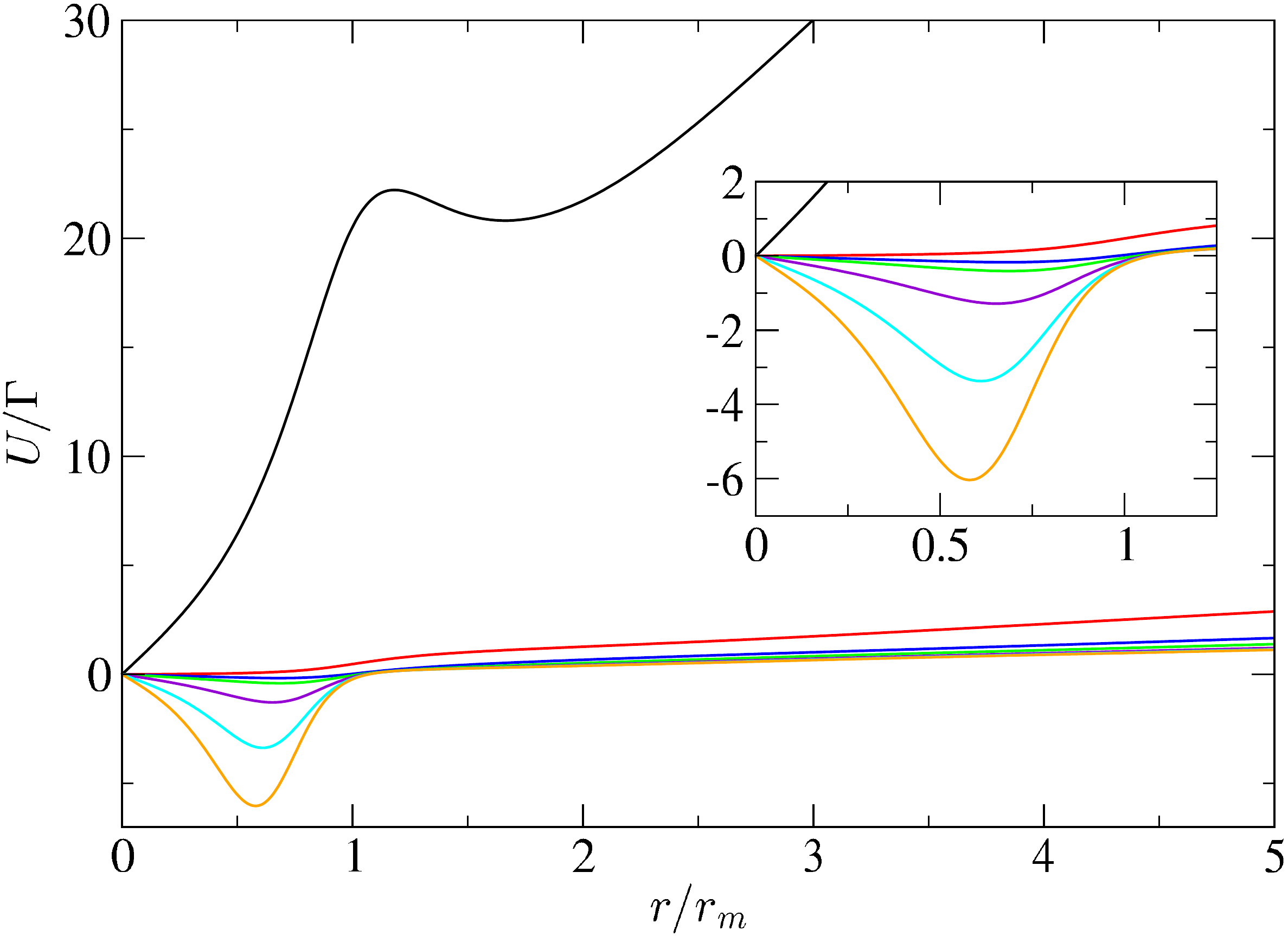}   
\includegraphics[width=0.8\columnwidth]{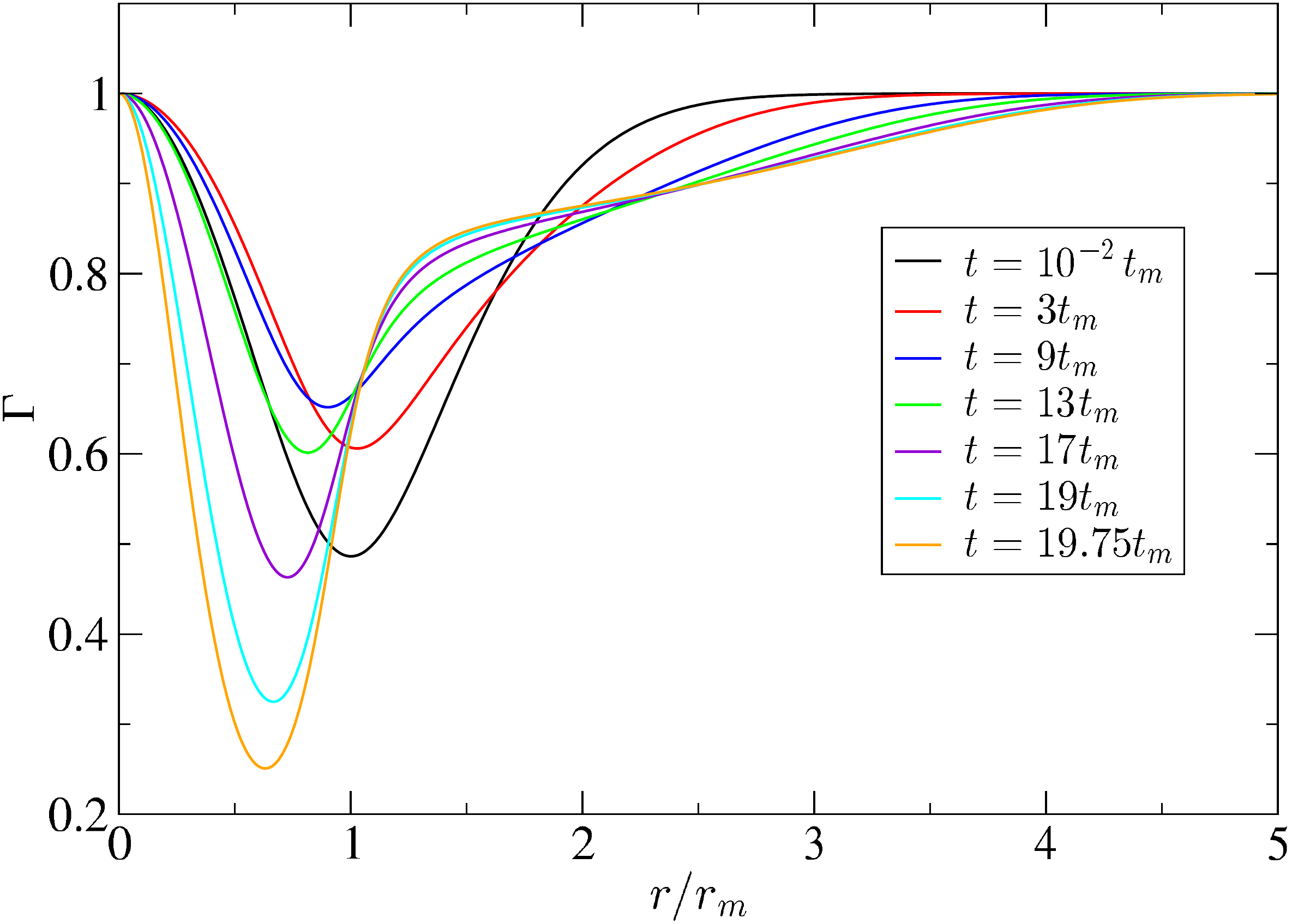}                                                                                                         
\caption{
Dynamical evolution of the different magnitudes at a given time $t$ for a supercritical perturbation in case of $q=1$ and $\delta=0.51$. We have taken $dt_{0} =10^{-3}$ and $N_{\rm cheb}=800$ in the simulation.}  
\label{fig:evolutioncritical}                                                                 
\end{center}                                                                     
\end{figure*}

\begin{figure*}[ht!]                                                                   
\begin{center}                                                                   
\includegraphics[width=0.8\columnwidth]{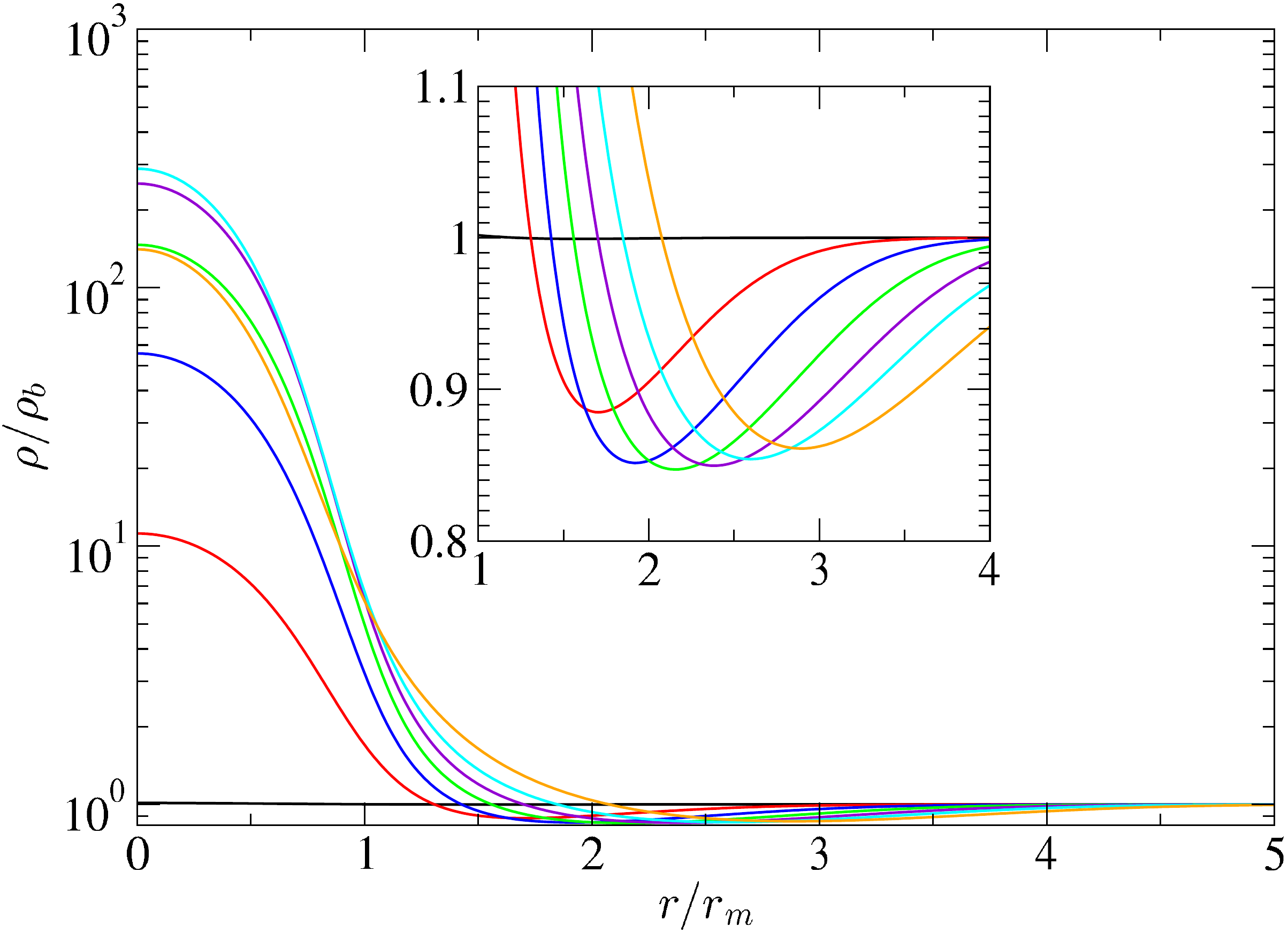}
\includegraphics[width=0.8\columnwidth]{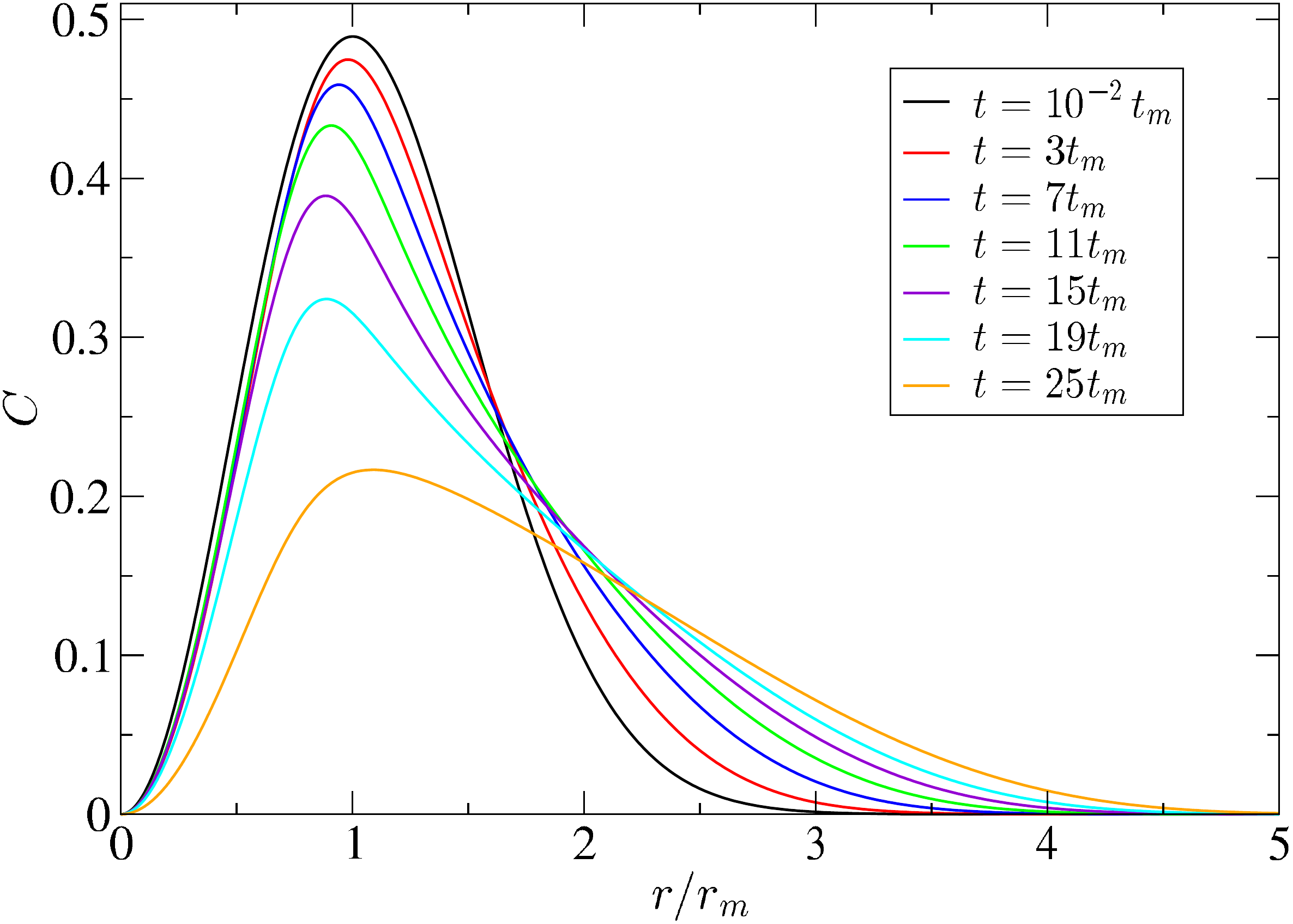}                            
\includegraphics[width=0.8\columnwidth]{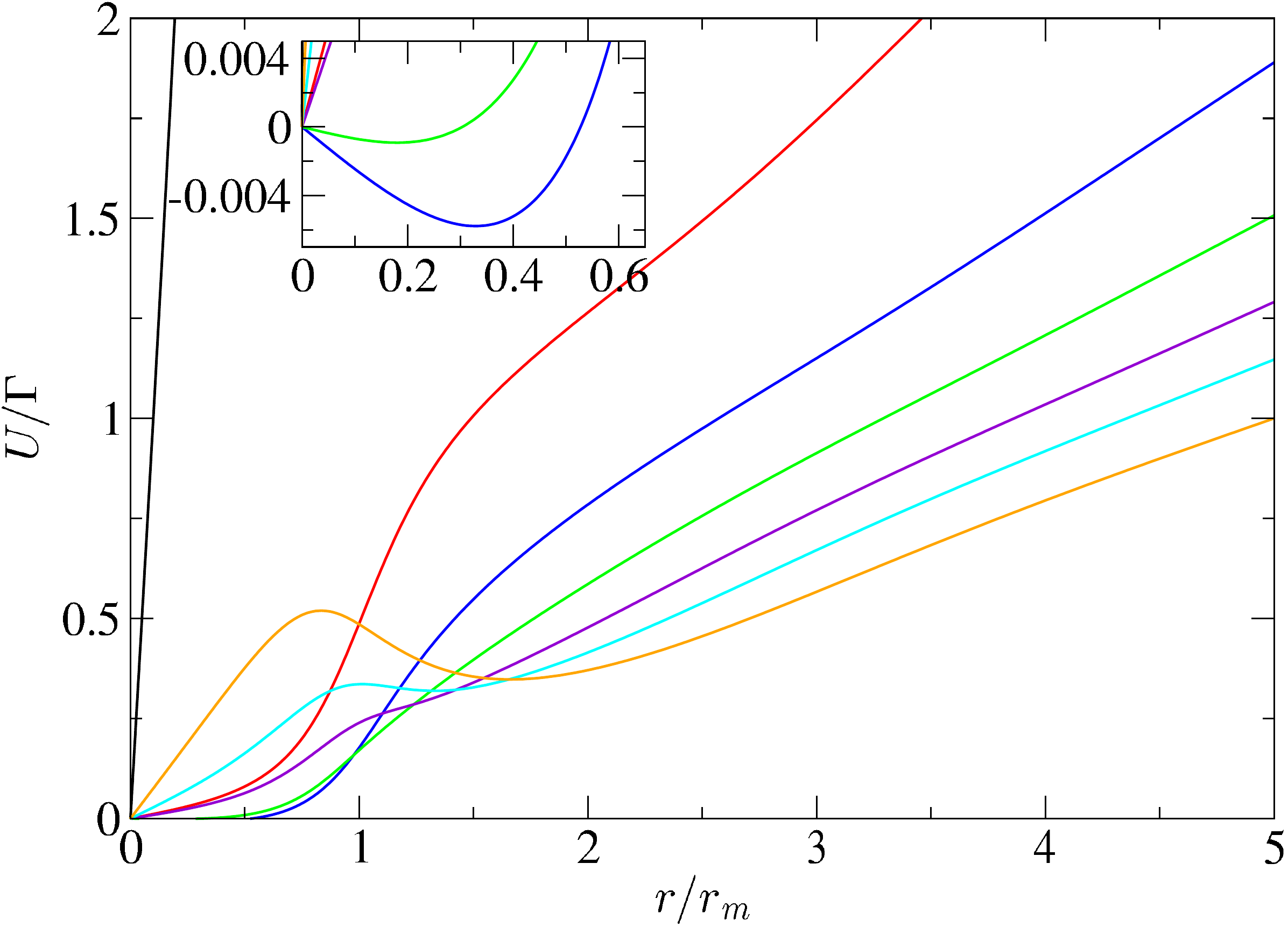}   
\includegraphics[width=0.8\columnwidth]{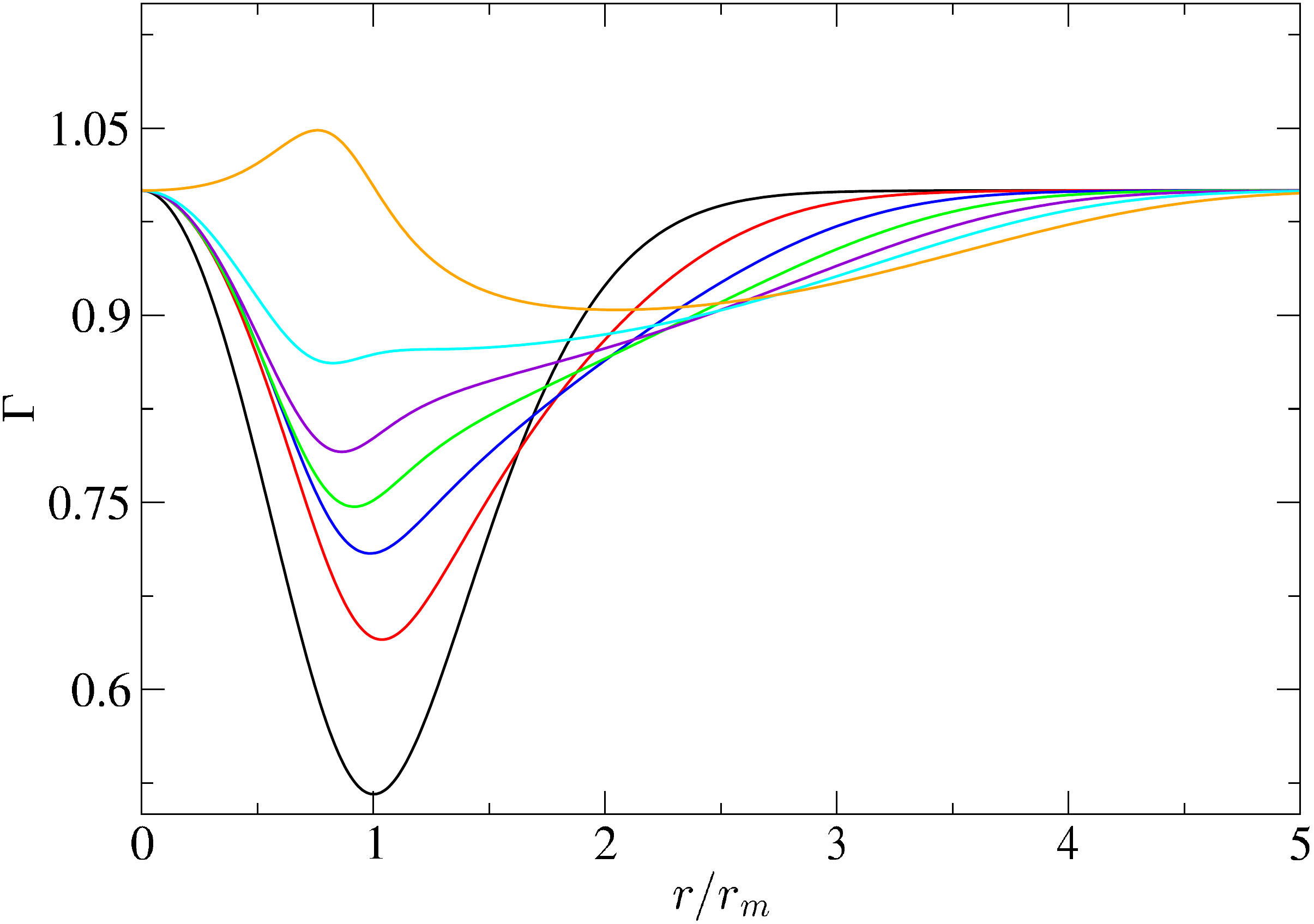}                                                                                                         
\caption{
Dynamical evolution of the different magnitudes at a given time $t$ for a subcritical perturbation in case of $q=1$ and $\delta=0.49$. We have taken $dt_{0} =10^{-3}$ and $N_{\rm cheb}=800$ in the simulation.}  
\label{fig:evolutionsubcritical}                                                                 
\end{center}                                                                     
\end{figure*}

\begin{figure*}[ht!]                                                                 
\begin{center}                                                                   
\includegraphics[width=0.8\columnwidth]{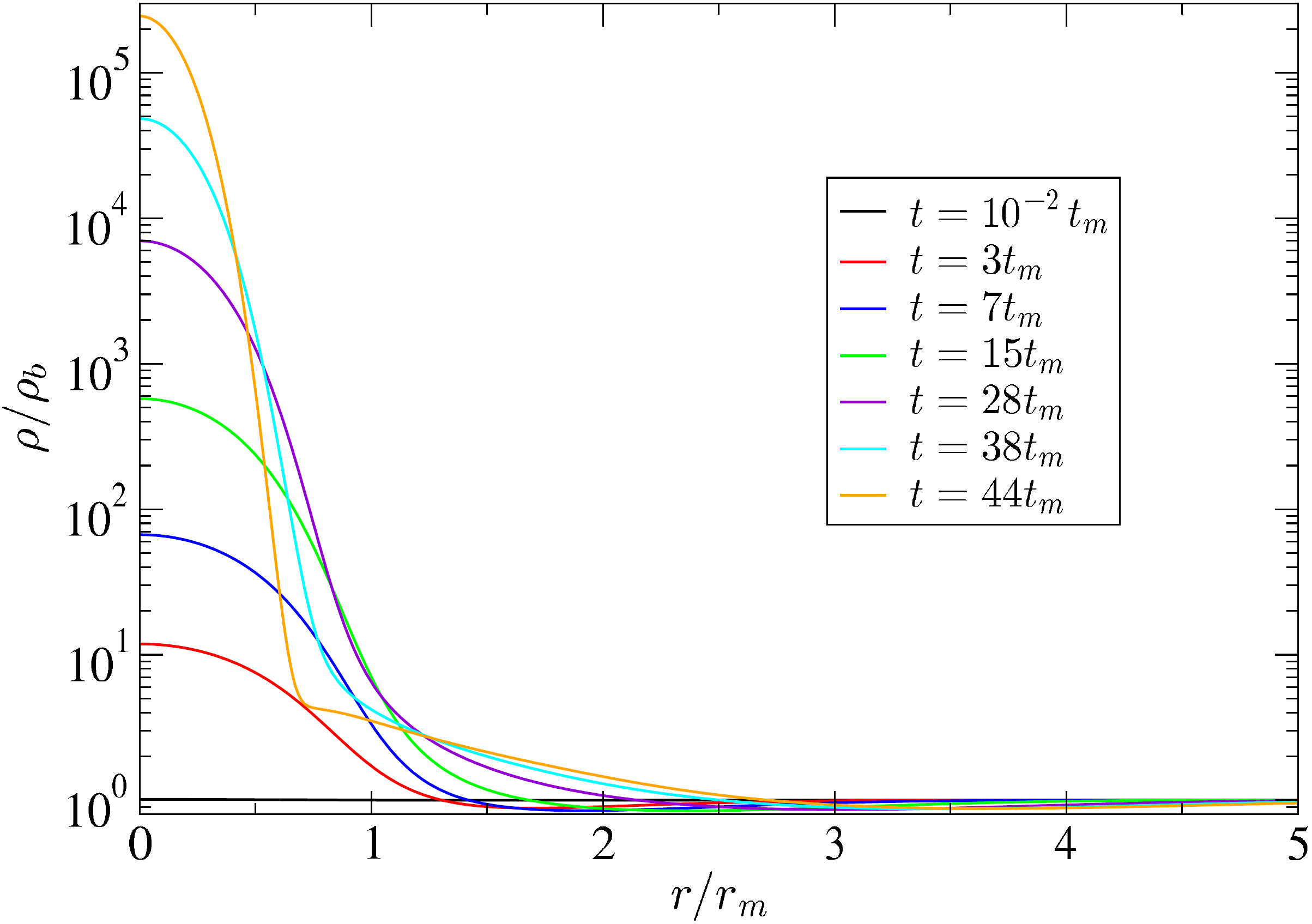}
\includegraphics[width=0.8\columnwidth]{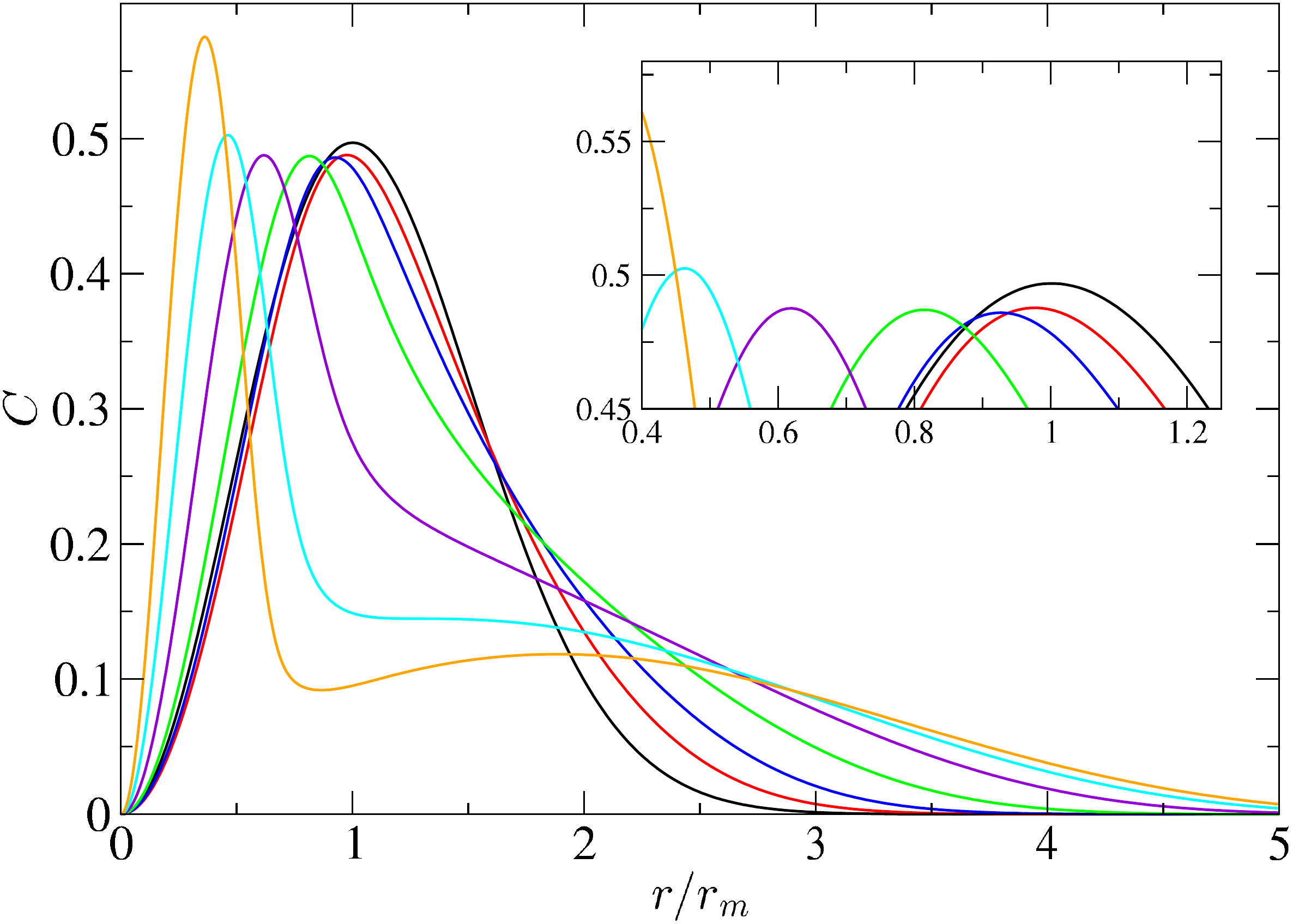}                            
\includegraphics[width=0.8\columnwidth]{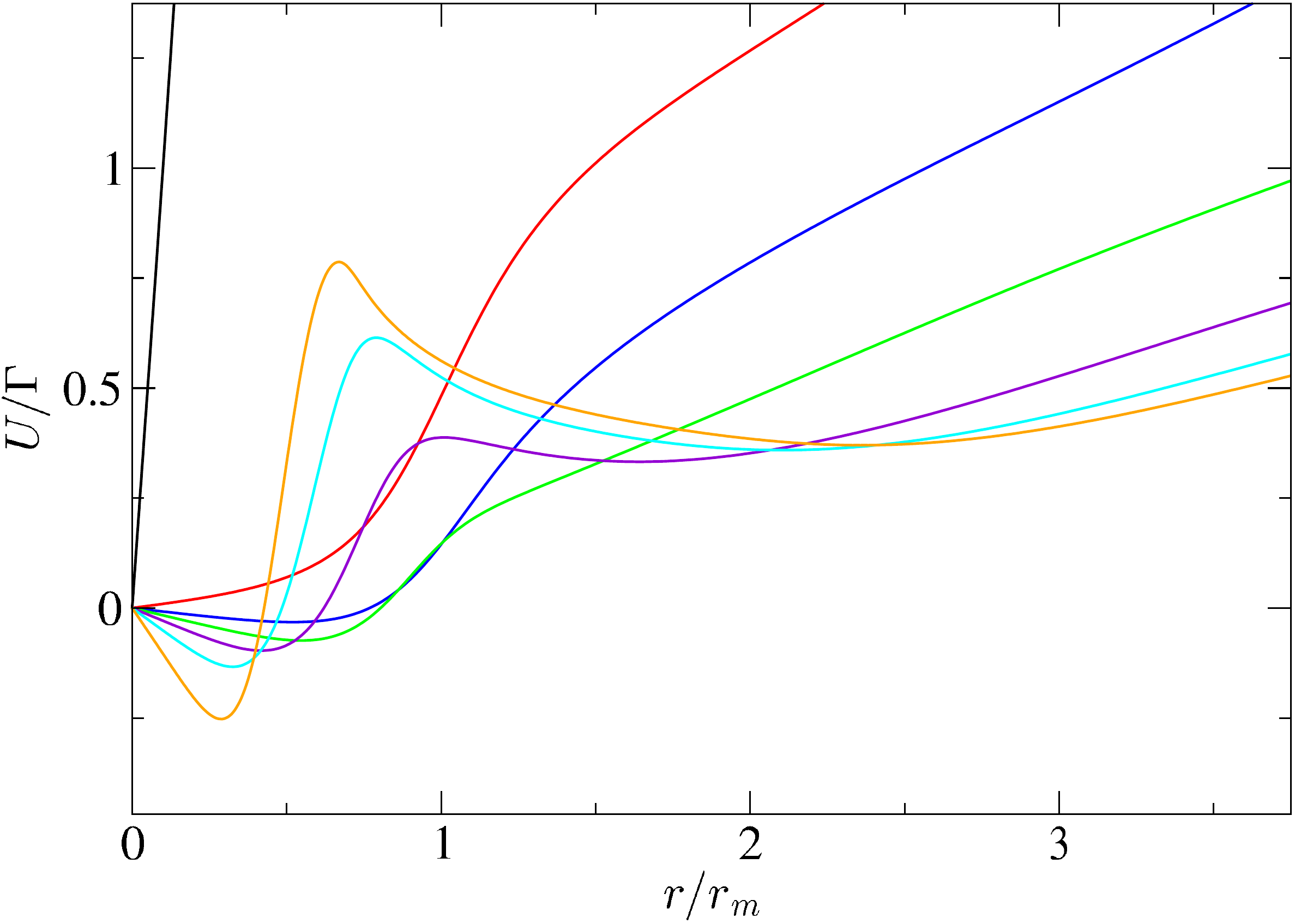}   
\includegraphics[width=0.8\columnwidth]{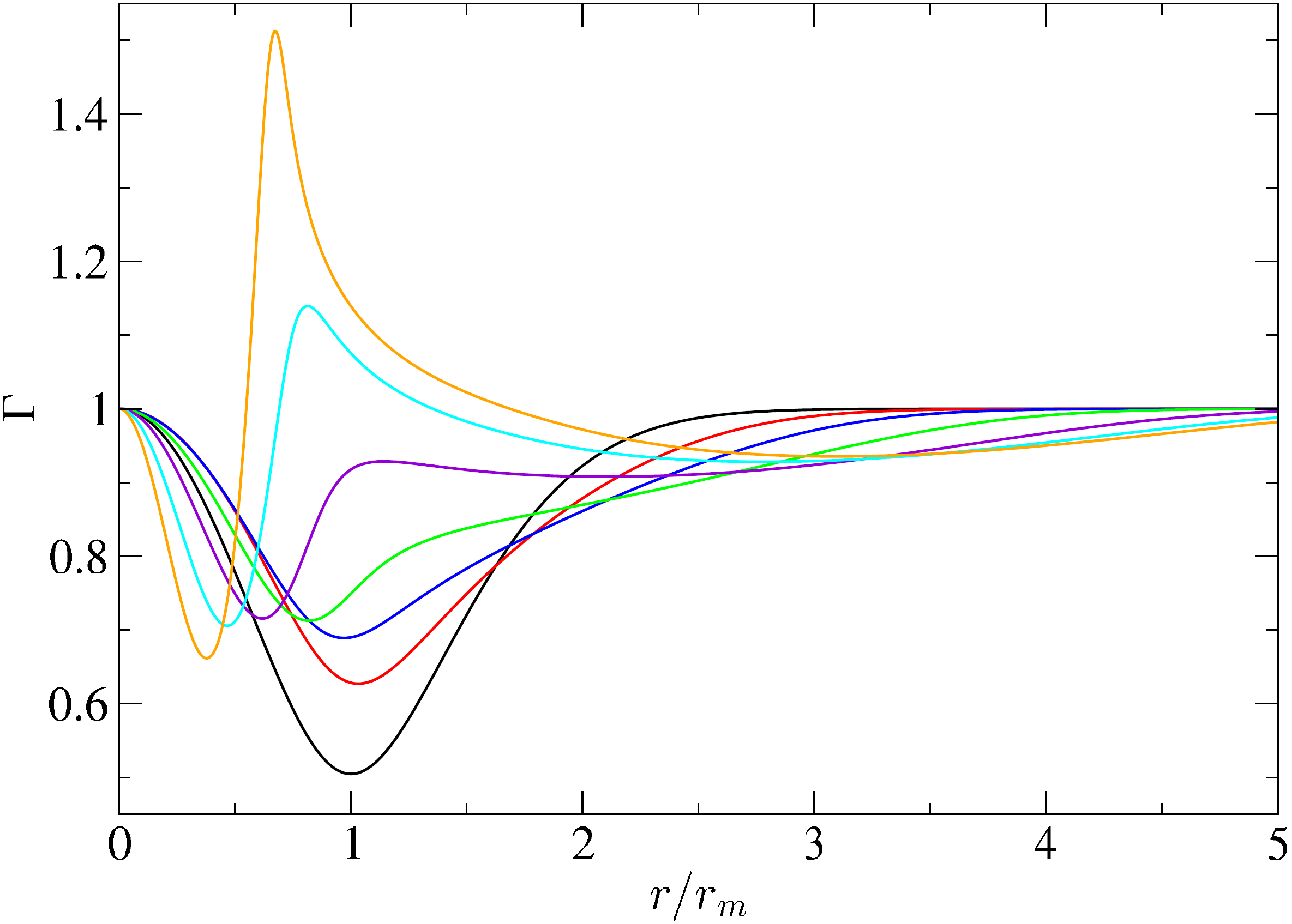}                                                                                                         
\caption{
Dynamical evolution of the different magnitudes at a given time $t$ for a perturbation with $\delta \approx \delta_{c}$ in case of $q=1$ with $\delta=0.49775$ and $\delta_{c} = 0.49774\pm 2\cdot 10^{-5}$. We have taken $dt_{0} =10^{-3}$ and $N_{\rm cheb}=800$ in the simulation.}  
\label{fig:evolutionclose}                                                                 
\end{center}                                                                     
\end{figure*}

In Figs.~\ref{fig:evolutioncritical},\ref{fig:evolutionsubcritical} and \ref{fig:evolutionclose} we see the evolution of the variables $\rho, \Gamma, U$ and $C$ for the Gaussian profile $q=1$ in the, respectively,  supercritical $(\delta > \delta_{c})$, subcritical ($\delta < \delta_{c}$) with $\abs{\delta-\delta_c} \gg O(10^{-3})$ and critical $\abs{\delta-\delta_c} \leqslant O(10^{-3})$ cases.

\begin{itemize}

\item In Fig.~\ref{fig:evolutioncritical} (the super-critical case) we see that the $C_{\rm max}$ grows during the evolution. From the same figure it is also evident the formation of two apparent horizons (where at the location of the horizons is satisfied that $2M/R=1$), as discussed in \cite{muscoelis}. The outer horizon moves outwards and the inner moves faster than the outer inwards. Once the inner horizon approaches the center of coordinates the simulation breaks due to the appearance of the singularity. 

In Fig.~\ref{fig:evolutionsubcritical} (the sub-critical case) $C_{\rm max}$ decreases continuously as the perturbation is diluted away due to the dominance of pressure gradients.

In Fig.~\ref{fig:evolutionclose} (the critical case) $C_{\rm max}$ first decreases and then bounces to re-increase again.

\item From the Figs.~\ref{fig:evolutioncritical}, \ref{fig:evolutionsubcritical} and \ref{fig:evolutionclose} we see that $\Gamma$ is not constant during the evolution. This implies, as it should, that the long wavelength approximation breaks down during the evolution.

\item In Fig.~\ref{fig:evolutioncritical} (super-critical case) we see that $U/\Gamma$ decreases quickly in time. Instead, in Fig.~\ref{fig:evolutionsubcritical} (sub-critical case) only a small negative value $U/\Gamma$ is reached for early times, and after that no negative values can be found, which means that the perturbation is dispersing avoiding the collapse. The most remarkable behavior is found in the critical case Fig.~\ref{fig:evolutionclose}.
Here the fluid splits into two parts, one going inwards (negative $U$) and one outwards (positive $U$) generating an under-dense region. This under-dense region re-attract the fluid with a net effect of a rarefaction and compression process which gets faster and faster. This is the reason why the code is not able to follow the evolution up to the final time BH formation.
\end{itemize}

\begin{figure}[ht!]
\centering
\includegraphics[width=1\linewidth]{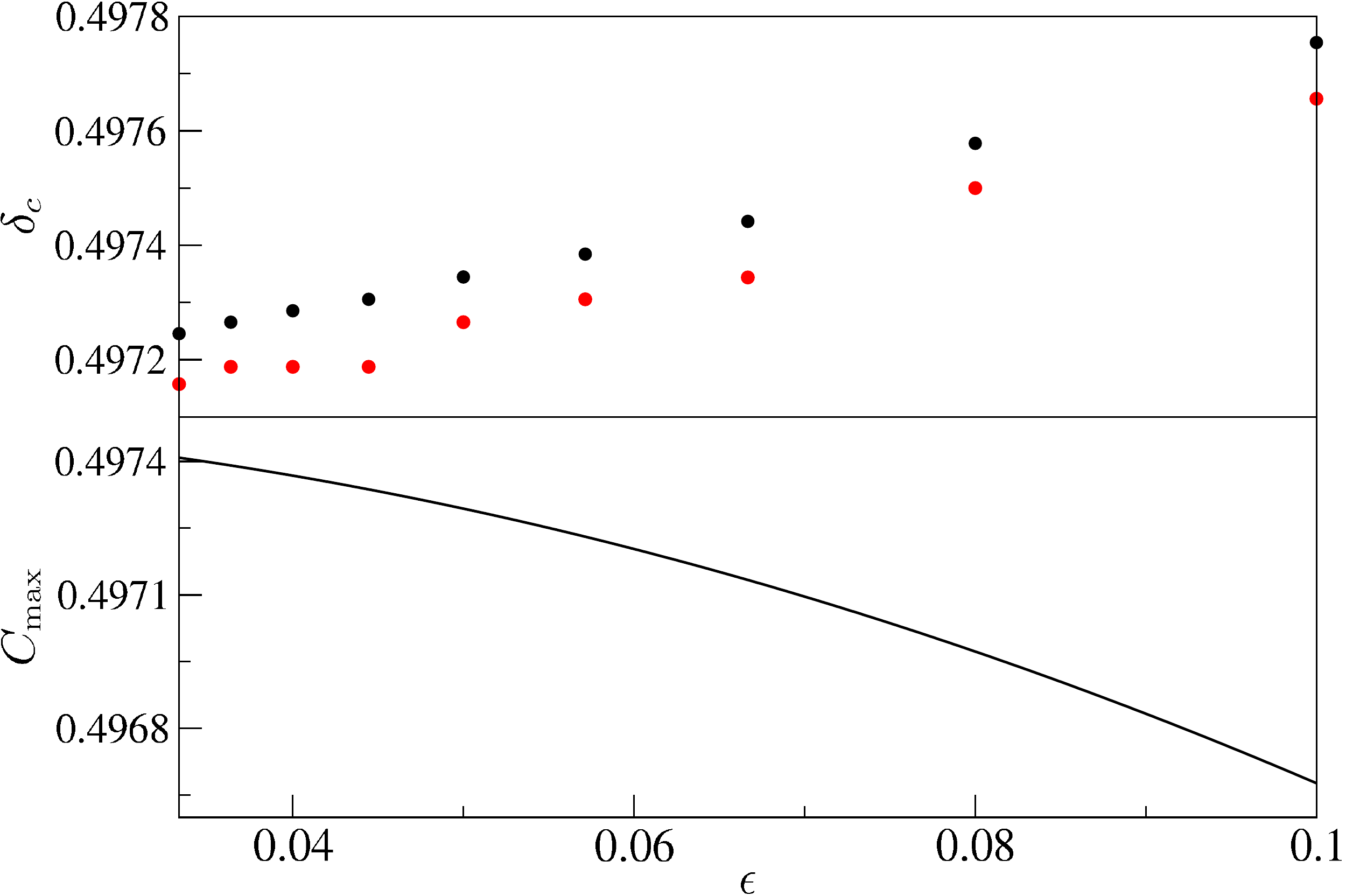} 
\caption{Top panel: threshold $\delta_{c}$ for the curvature Gaussian profile for different values of $\epsilon$, taking $N_{\rm cheb}=400$ and $dt_{0}=10^{-3}$. Black points are $\delta_{c,\rm yes}$ and red points $\delta_{c,\rm no}$. Bottom panel: $C_{\rm max}$ in terms of $\epsilon$ computed with Eq.(\ref{compactionfunction}).}
\label{fig:longwave}
\end{figure}

Let us finally remark something about the long wavelength approximation. As can be seen in Fig.~\ref{fig:longwave} the threshold $\delta_{c}$ (as well as $C_{\rm max}$) has some small dependence in terms of $\epsilon$. It is obvious that the difference between the asymptotic critical value and the one numerically found grows with $\epsilon$. Thus, a physical limitation (not numerical) on the resolution of $\delta_c$ of $O(10^{-3})$ is already present, due to the use of the long wavelength approximation to build the initial conditions.

\subsection{Power-spectrum profile}

In this section, we aim to provide a test of the stability of our code for profiles that differ from the ones studied before in Eq.(\ref{profile}). The main difference are under- and over-density oscillations away from the peak of the curvature. 

The profiles used here are sub-classes of the mean profiles obtained with the procedure outlined in \cite{germaniprl} by broken power spectrums of the form 

\begin{equation}
P(k) = 
\begin{cases} 
      0 & k < k_{p} \\
      P_{0} \left(\frac{k}{k_{p}}\right)^{-n} & k \geq k_{p}\ , \\
   \end{cases}
\label{eq:powerspectrumtemplate}
\end{equation}

which are relevant for cosmological applications \cite{germani-vicente}. In particular, we shall only consider the convergent cases of $n\geq 0$. In Eq.(\ref{eq:powerspectrumtemplate}) $k_{p}$ is the wavelength of the peak. After a straightforward computation, one finds that the mean curvature is
\begin{figure}[ht!]
\includegraphics[width=1\linewidth]{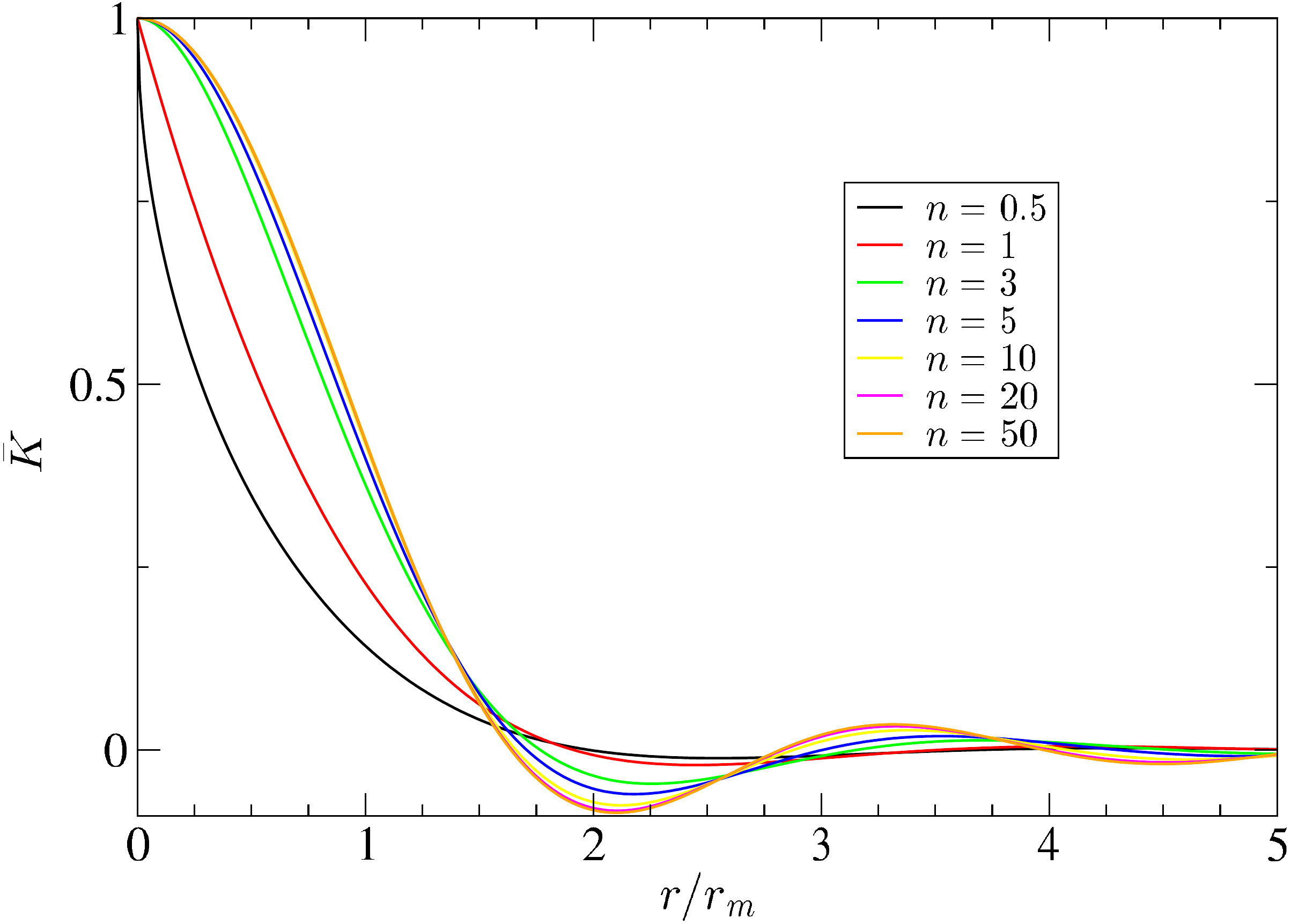} 
\caption{Curvature profile $\bar{K}(r)$ in terms of $n$ using Eq.(\ref{eq:powercurvature}).}
\label{fig:spectrum}
\end{figure}

\begin{align}
\label{eq:powercurvature}
\bar{K}(r)  &= \frac{3 n}{2 (k_{p} r)^3}\left[-k_{p}r \left\{ E_{3+n}(-ik_{p}r)+E_{3+n}(ik_{p}r)\right\} \right. \nonumber \\
&+ \left. i\left\{E_{4+n}(ik_{p}r)-E_{4+n}(-ik_{p}r) \right\} \right],
\end{align}

where
\begin{equation}
E_{n}(x) = \int_{1}^{\infty}\frac{e^{-x t}}{t^{n}}dt.
\end{equation}
 
\begin{figure}[ht!]
\centering
\includegraphics[width=1\linewidth]{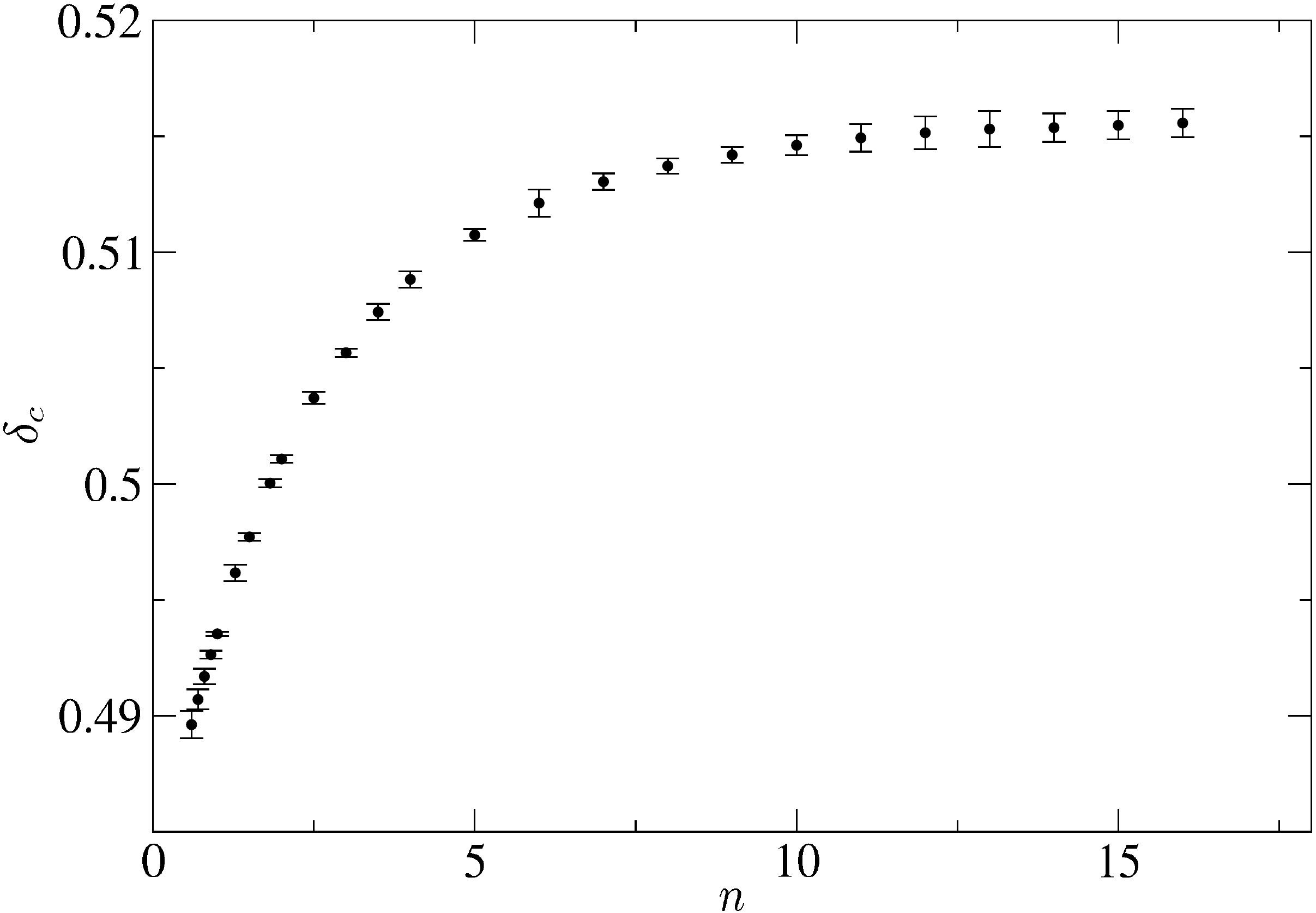} 
\caption{Values of $\delta_{c}$ for different values of $n$ for the curvature profile of Eq.(\ref{eq:powercurvature}). Simulations done with $N_{\rm cheb} \approx 700$ and $dt_{0}=10^{-3}$.}
\label{fig:thresholdnongausian}
\end{figure}

From a given value of $r_{m}$ and $n$, we get the correspondent value of $k_{p}$ solving numerically Eq.(\ref{cmax}). An important difference from these profile with respect to the ones studied before is that here we needed to consider a larger number of $N_{\rm cheb}$ in order to capture the oscillations of the curvature. Finally, in Fig.~\ref{fig:thresholdnongausian} are shown the thresholds obtained for different values of $n$. 

\begin{figure}[ht!]
\centering
\includegraphics[width=1\linewidth]{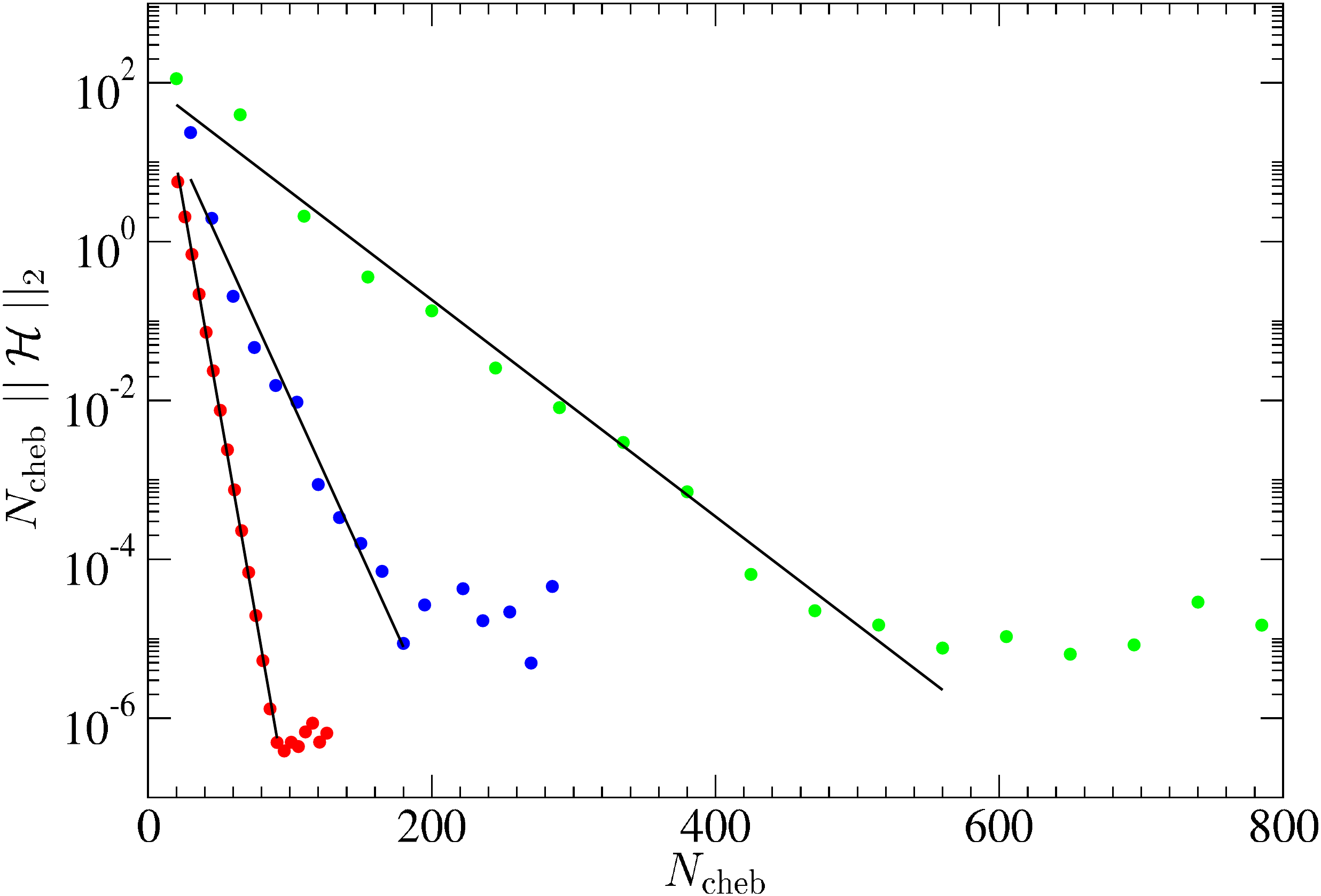} 
\caption{Spectral convergence for different curvature profiles. Red points corresponds to the profile of Eq.(\ref{profile}) with $q=1$, green points corresponds to $q=5$ and blue points to the profile of Eq.(\ref{eq:powercurvature}) with $n=15$. The black solid line is the exponential fit $\sim e^{-\alpha N_{\rm cheb}}$ with $\alpha \approx 0.23, 0.031, 0.092$ respectively for the cases quoted before.}
\label{fig:spectral_acuracy}
\end{figure}

Finally, we have tested the spectral convergence of the profiles considered in terms of the Hamiltonian constraint, the results can be seen in Fig.~\ref{fig:spectral_acuracy}.

\section{Mass spectrum}

It is known that for $\bar\delta(r_m)$ close to the critical value $\delta_c$ the mass of the black hole follows the following scaling law \cite{musco2009,Niemeyer2,hawke2002}
\begin{equation}
M_{BH} = M_{H} {\cal K} (\delta-\delta_{c})^{\gamma},
\label{eq:scaling}
\end{equation}
where $\gamma \approx 0.36$ in radiation. In Eq.(\ref{eq:scaling}) the constant ${\cal K}$ is a correction factor due to the choice of the reference mass $M_H\equiv 1/2 H(t_m)$, where the Hubble scale has been calculated at the time $r_m H(t_m)a(t_m)=1$. The scaling law starts to deviate at $(\delta-\delta_{c}) \gtrsim 2 \cdot 10^{-2}$, \cite{musco2009}. 

To test our code, in this section we will numerically obtain the constant $\cal{K}$, for a Gaussian profile. Moreover, in the cosmological context, one needs the value of  ${\cal K}$ to estimate the PBH abundances \cite{germaniprl}. 

Previous numerical computations were performed in the region up to $(\delta-\delta_{c}) \approx 10^{-1.2}$. We will show in the following, for the first time, the mass range for large values of $\bar\delta(r_m)$ up to the maximal value $2/3$.

The way we will find the mass spectrum is by the implementation of an excision technique \cite{bookNR} which avoids the region of large curvatures in the Misner-Sharp evolution where the code would break.

The key idea of excision is that the evolution of matter inside the horizon cannot affect the physics outside. The excisions follow the motion of the apparent horizon. The implementation of this technique is straightforward using spectral method, in contrast with finite differences~\citep{Teukolsky}, since the derivative at the excision boundary (that we have to define when we cut part of the computational domain) is computed without taking into account points that lies inside the inner boundary (in finite differences it is necessary to interpolate).

Unfortunately, the excision technique cannot be used until the formation of the black hole. This is due to the fact that the velocity of the outer horizon is too small and the initial resolution is not enough to follow the change in apparent horizon. Of course this can be solved with an implementation of some kind of AMR for spectral methods, like junctions of Chebyshev grids. We will however follow here another (semi-analytical) direction.

To estimate the final mass of the PBH, we have used the Zeldovich-Novikov formula Eq.(\ref{eq:ZN_formula}), which assumes Bondi accretion \cite{acreation1}. It is important to highlight that this is not applicable at the moment of formation of the horizon, since it neglects the cosmological expansion \cite{Carr_acreation}, but we can apply from sufficiently late times after the formation of the PBH considering an effective constant accretion rate $F$ \cite{acreation2,acreation3}. This approximation was already employed in the context of PBH formation from domain walls in \cite{deng}.

In particular, at the final stage of the BH formation, the mass accretion  follows the law 
\begin{equation}
\label{eq:ZN_formula}
\frac{dM}{dt} = 4 \pi F R^2_{\rm BH} \rho_{b}(t)\ .
\end{equation}

$F$  is usually numerically found to be of order $O(1)$. By the condition of apparent horizon $R_{\rm BH}=2M_{\rm BH}$, the previous equation is solved as:
\begin{equation}
\label{eq:acretationformula}
M_{\rm BH}(t) = \frac{1}{\frac{1}{M_{a}}+\frac{3}{2}F\left(\frac{1}{t}-\frac{1}{t_{a} }\right)}\ ,
\end{equation}
where $M_a$ is the initial mass when the asymptotic approximation is used at the time $t_a$ .

We will find $F$ by fitting the numerical evolution of the mass via the excision method. Once found it, the PBH mass will be inferred as the asymptotic mass at $t\rightarrow \infty$, i.e.

\begin{equation}
\label{eq:massfinalfinal}
M_{\rm BH}(t \rightarrow \infty) = \left(\frac{1}{M_{a}}-\frac{3 F}{2 t_{a}}\right)^{-1} \ .
\end{equation}

\subsection{Excision technique}

The main idea of the excision technique implemented here is to dynamically remove part of the computational domain within the horizon, that would otherwise develop large gradients and eventually break down the simulation. 

To do that, we have defined two parameters, $\Delta r$ and $dr$. $\Delta r$ is the separation between the excision boundary and the apparent horizon that we set after each redefinition of the excision surface. $dr$ is the maximum allowed displacement of the apparent horizon before we redefine the excision surface. We consider always that $\Delta r>dr$. 

We locate the position of the apparent horizon (defined as $2M(r,t)/R(r,t)=1$) after each time step using a cubic spline interpolation (we have checked that the difference in $M(r,t)$ taking a quadratic spline interpolation are $O(0.01\%)$). 

Specifically, the exact procedure we have used is the following:

At the time when $C_{\rm max} \approx 1.2$ (the result is not affected by the exact choice as long as $C_{\rm max} \approx O(1)$), we remove part of the computation domain creating an excision surface close to the apparent horizon whose separation with the excision boundary is precisely given  by $\Delta r$. After that, the system is evolved as usual in the new Chebyshev grid with the new domain (the Chebyshev differentiation matrix has to be redefined as well). Once the apparent horizon has displaced a distance greater than $dr$, we redefine a new excision surface close to the new location of the apparent horizon, again with the same separation $\Delta r$. We repeat this process continuously. 

The values of $\Delta r$ and $dr$ are slightly reduced in time when is needed. This is particularly important for the smallest values of $\delta-\delta_{c}$. To do that, when a simulation is going to break down due to large gradients, we return to a "safe point", reducing $\Delta r$ and $dr$. After that, we proceed with the usual way. 

The values that we have considered are $\Delta r \approx 2 dr \approx O(10^{-2})$. $\Delta r$ and $dr$ can not be taken arbitrarily small, due to the limitation of the resolution given by the Chebyshev grid. An AMR can solve this, but the current implementation worked already well for our purposes.

Although we didn't apply boundary conditions at the excision surface, (in comparison with $r=0$) we found that freezing the value of $\rho'$ at the excision surface, after each redefinition of the boundary, increases the stability of the procedure without changing the results.

For the computation of the excision we have taken at least $N_{\rm cheb}= 1000$, to increase the resolution and be able to make the excision sequentially.

\subsection{Numerical results}

\begin{figure}[ht!]
\centering
\includegraphics[width=1\linewidth]{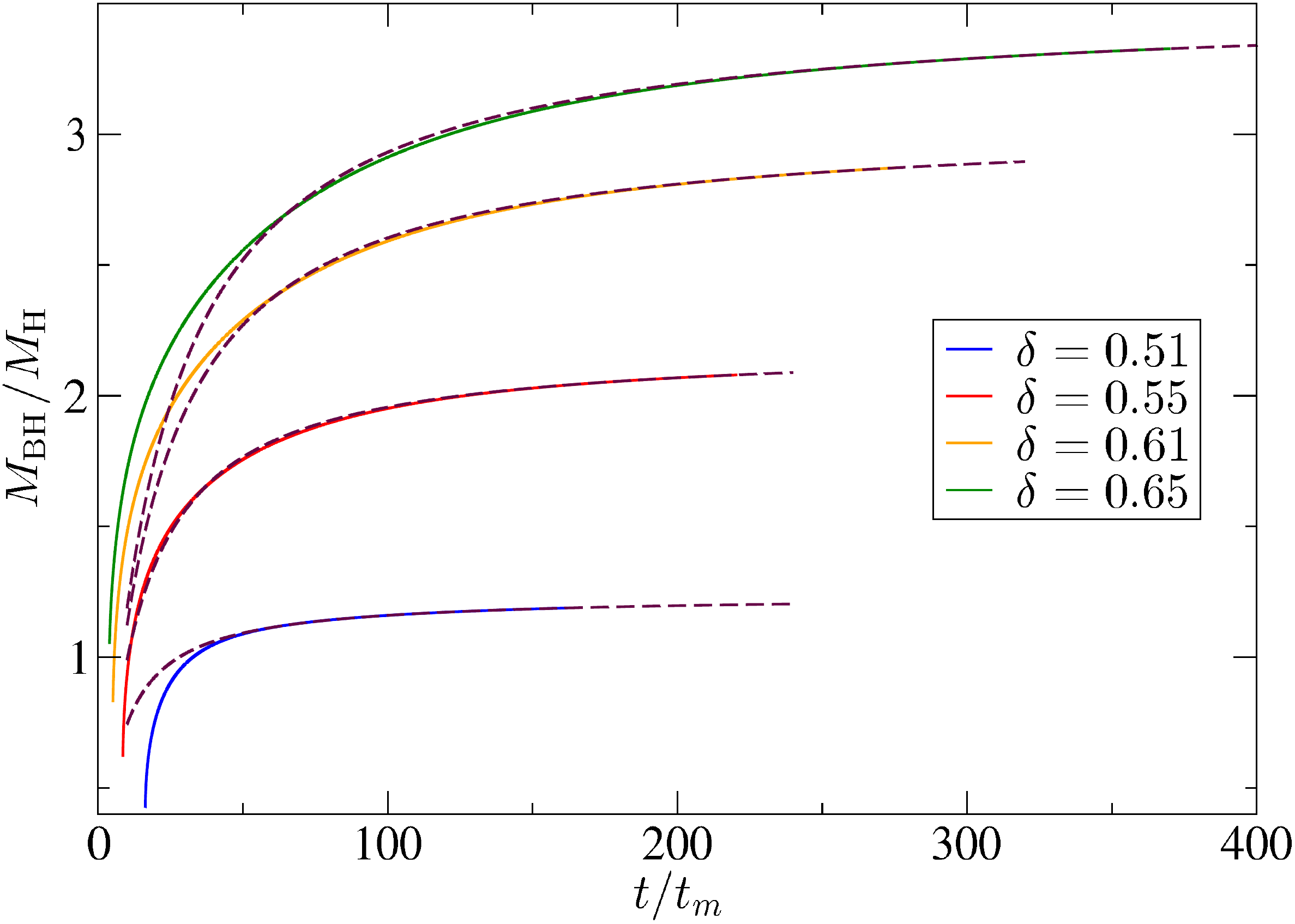} 
\caption{Mass of the BH in time after the formation of the apparent horizon for different values of $\delta_{c}$. The dashed line corresponds to the analytical fit with Eq.(\ref{eq:acretationformula}).}
\label{fig:mass-evolution}
\end{figure}

The evolution of the black hole mass in time $M_{BH}(t)$ can be seen in Fig.~\ref{fig:mass-evolution}.

\begin{figure}[ht!]
\centering
\includegraphics[width=1\linewidth]{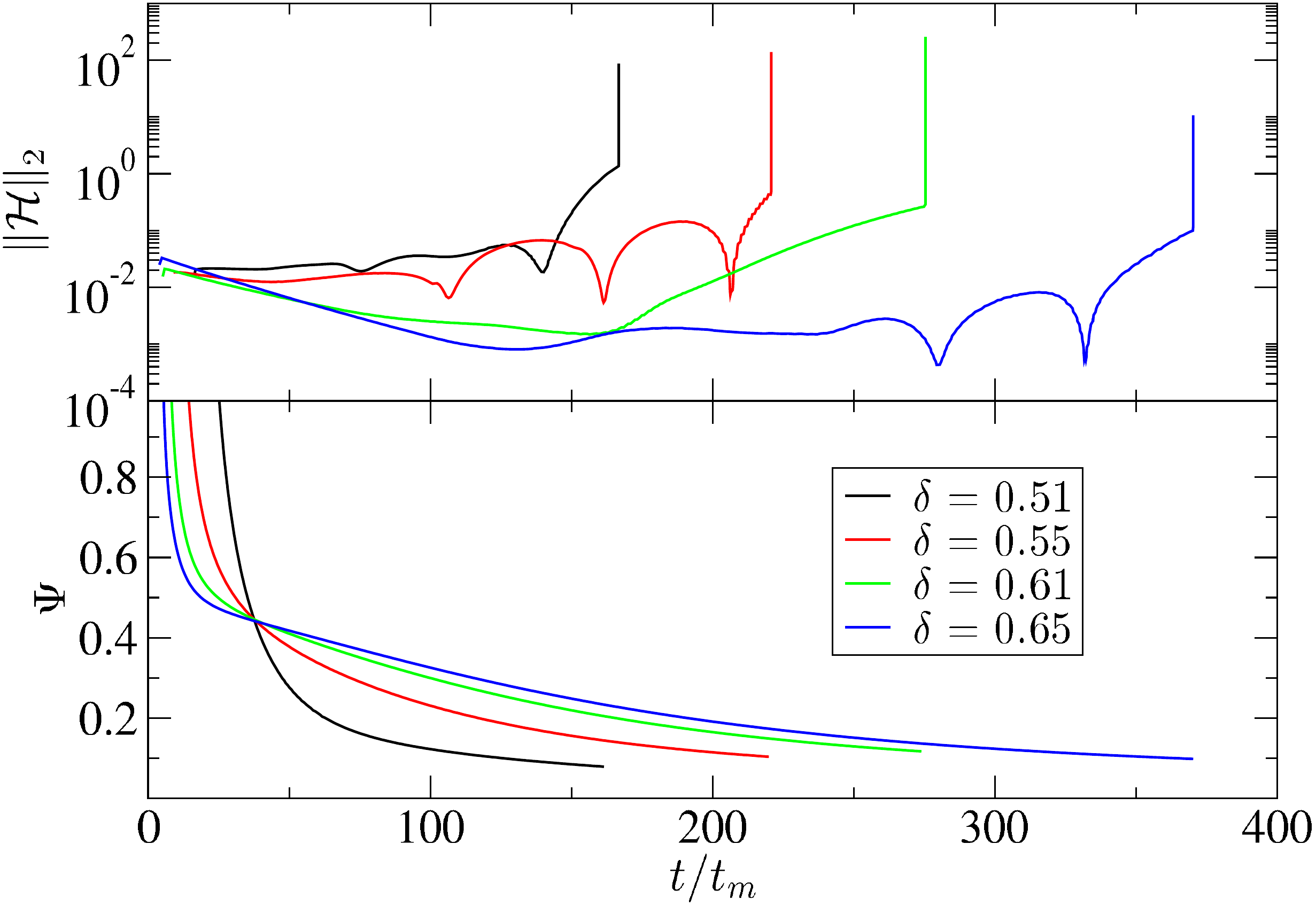} 
\caption{Top panel: Hamiltonian constraint during the excision procedure for different values of $\delta$. Bottom panel: Evolution of $\Psi$ in time. The crossing point is around $t/t_{m}\approx 37.5$.}
\label{fig:constraint-excision}
\end{figure}

In order to check when the approximation of Eq.(\ref{eq:acretationformula}) is valid, we have computed the ratio of the increment of the black hole mass respect the Hubble scale $\Psi = \dot{M}/H M$, which is expected to be $\Psi<1$ when the evolution satisfy this regime. We have made a non-linear fit in the Eq.(\ref{eq:acretationformula}) to get the parameters $t_{a}$, $M_{a}$ and $F$ to estimate the mass of the black hole. The range of numerical values that we use to make the fit are those which fulfill $\Psi \lesssim 0.1$, which works well for our purposes. We have checked that the Hamiltonian constraint is fulfill until late time, when the simulation breaks, Fig.~\ref{fig:constraint-excision}. Nevertheless, we have tested that the evolution of the mass is not affected by the violation of the constraint. The results can be found in  Fig.~\ref{fig:constraint-excision}. Interestingly, we see a crossing for different evolution of $\Psi$ at a given time $t^{*}$.

The values of $F$ that we get goes from $F\in [3.5,3.75]$ increasing the value of $\delta$. This is consistent with the one reported in \cite{deng} where a value of $F \approx 3.8$ was got for large black holes, although the mechanism of PBH formation is different. We have checked always that the fit performed is accurate, getting a variance of $\sigma_{\rm max} \approx 10^{-2.5}$. The standard deviation $s_{d}$ of the parameters are $s_{d}(t_{a}) \approx 10^{-9}$, $s_{d}(M_{a}) \approx 10^{-5}$ and $s_{d}(F) = 10^{-5}$.

\begin{figure}[ht!]
\centering
\includegraphics[width=1\linewidth]{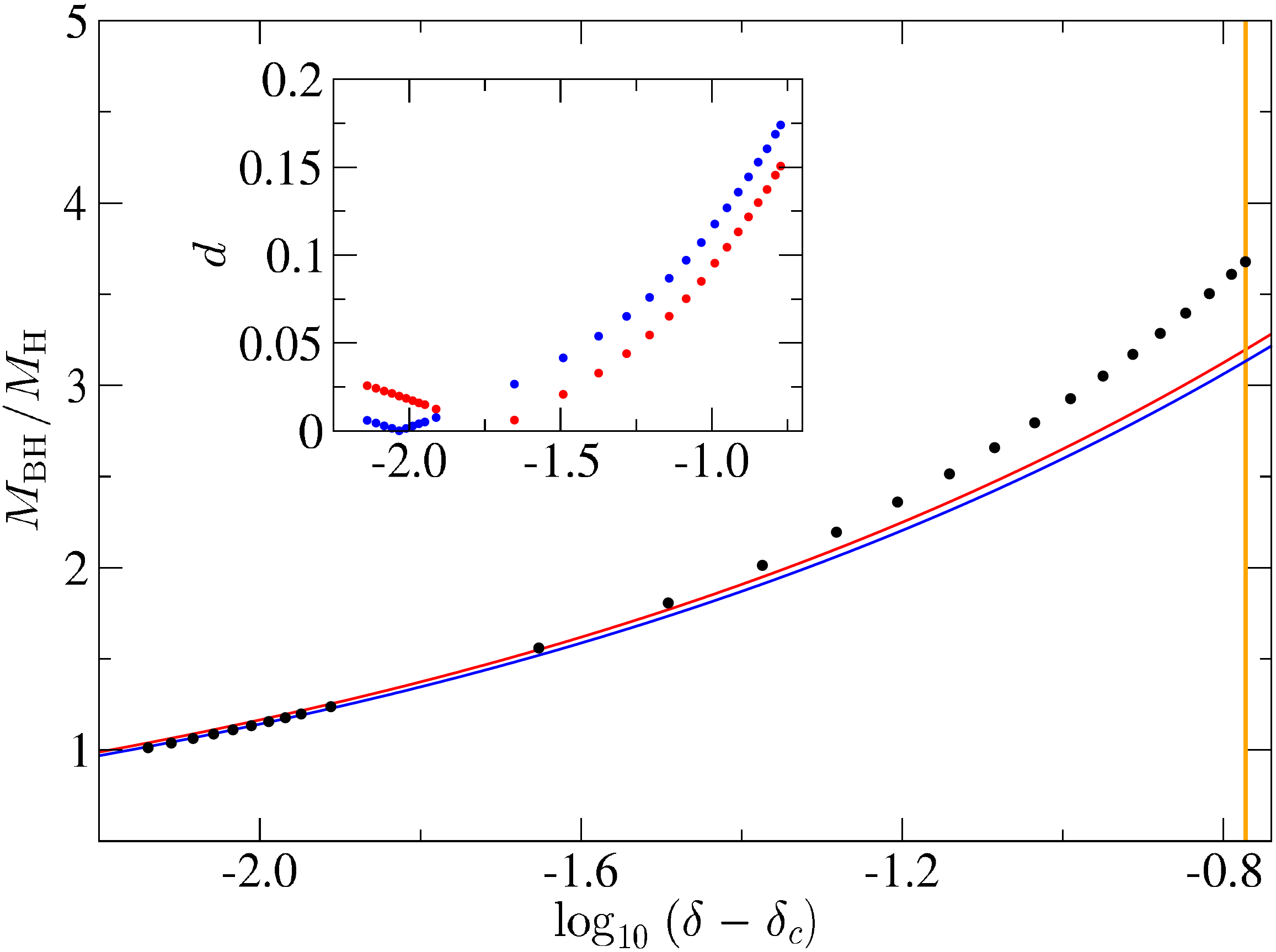} 
\caption{Values of $M_{\rm BH}/M_{H}$ in terms of $(\delta-\delta_{c})$. The solid red line corresponds to the scaling law behaviour with $\gamma=0.357$, $\delta_{c}=0.49774$ for ${\cal K}=6.03$ and the blue solid line with the numerical value for ${\cal K}=5.91$. Dark points are the values got from the fitting of Eq.(\ref{eq:massfinalfinal}). The subplot represents the absolute value of the relative deviation $d$ respect the numerical values and the ones coming from the scaling law. The orange vertical line is the value $\delta_{\rm max}=2/3$.}
\label{fit:scalingfit}
\end{figure}

We have used the values of $M_{BH}$ in the range of $\delta \in [0.505,0.51]$ to estimate the value of $\cal{K}$ from the scaling law, taking into account that $\delta_{c} = 0.49774$ and $\gamma =0.357$. The values of ${\cal K}$ in this domain of $\delta$ are ${\cal K} \in [5.87,5.96]$, making an average we get ${\cal K} = 5.91$. This values differs in $1.9\%$ from the value quoted in the literature with ${\cal K} = 6.03$. The values of $M_{\rm BH}$ in terms of $\delta$ can be found in Fig.~\ref{fit:scalingfit}.

Finally, for the first time we present the values of $M_{\rm BH}$ for large values of $\delta$ until $\delta_{\rm max} = 2/3$. We observe that the scaling law deviates at the higher end of in the $\delta$ range up to $O(15\%)$, as can be seen in the subplot of Fig.~\ref{fit:scalingfit}. For this particular case we obtain that the maximum allowed mass of the black hole is $M_{\rm max(BH)} \approx 3.7 M_{\rm H}$. Is expected that this deviation is not going to significantly affect the PBH abundances due to the rarity of such perturbations.

\section{Conclusions}

We have performed numerical simulations of PBHs formations using Pseudo-spectral methods instead of the extensively used Lagrangian hydrodynamic formalism based on \cite{maywhite,baumgarte}. We have been able to obtain the threshold $\delta_{c}$ of different curvature profiles with up to an accuracy of $O(10^{-5})$, which match with the ones quoted in the literature \cite{musco2018}. Our method is simple and efficient and allows to estimate the thresholds with enough accuracy for cosmological applications, where an accuracy of $O(10^{-2})$ in $\delta_{c}$ is required \cite{germaniprl}.

In our simulations we have used an excision technique to remove the singularity from the computational domain. To get the mass of the black hole, we have employed a semi-analytical formalism given by Eq.(\ref{eq:acretationformula}), which leads a deviation of $O(2\%)$ in the determination of the black hole mass with respect to the values quoted in the literature, in the scaling law regime. Moreover, for the first time we were able to give the values of the black hole mass for large initial amplitudes, finding a deviation of $O(15\%)$ at the largest value $\delta_{\rm max}=2/3$ with respect to Eq(\ref{eq:scaling}).

Our code is an independent test of the correctness of the thresholds found earlier in the literature. The present algorithm can be used in the cosmological context of PBH formation in a FRW background, as it has been already successfully done in \cite{RGE,garrigavicentejudithescriva}. Moreover, our method could be the way to solve a multidimensional collapse because the standard implementation of the hydrodynamical methods seems to fail \cite{rezolla}. However we leave it for future research.

\begin{acknowledgments}

I would like to thank Cristiano Germani for many suggestions and for checking in detail this draft. Also I would like to thanks Vicente Atal, Jaume Garriga, Ian Hawke and Ilia Musco for interesting discussions. A special thank to Carsten Gundlach for discussions and hospitality in Southampton  University, as well as for useful comments about the first version of the manuscript. Thanks as well to the anonymous referees for the suggestions to improve the submitted draft. Finally I wish to thank Javier G. Subils for suggesting the use of spectral methods and discussions on it. I am partially supported by the national FPA2016-76005-C2-2-P grants and supported by the Spanish MECD fellowship FPU15/03583.

\end{acknowledgments}


\bibliography{refs.bib}

\end{document}